\documentclass[11pt, a4paper]{article}
\usepackage{graphicx}
\usepackage{natbib}
\usepackage{txfonts}
\usepackage{amssymb}
\usepackage{url}
\usepackage[a4paper,total={17cm,25cm}]{geometry}
\begin{document}
\title{Wide-field Polarization Imaging and Numerical Modeling of the Coma and Tail of Comet C/2023~A3 (Tsuchinshan-ATLAS)}
\author{M. Arnaut$^1$ \and C. W\"ohler$^1$ \and P. Halder$^2$ \and G. Ahuja$^3$ \and S. Ganesh$^3$ \and M. Bhatt$^3$}
\date{%
$^1$Image Analysis Group, TU~Dortmund University, Otto-Hahn-Str.~4, 44227 Dortmund, Germany\\
$^2$Dept.\ of Physics and Astronomy, University of Nebraska-Lincoln, Lincoln, NE, 68588-0299, USA\\
$^3$Physical Research Laboratory, Ahmedabad, 380009, India\\[2ex]
{\em Preprint, \today}
}
\maketitle
\abstract{Imaging polarimetry enables the spatially resolved investigation of cometary dust properties across different morphological structures. While cometary comae have been studied thoroughly in the pertinent literature, cometary tails have remained less explored. Comparing these regions can reveal differences in the size, structure, and composition of their dust. The goal of this study is to examine the size, structure and composition of the dust particles in the coma and in particular in the tail of the bright comet C/2023~A3 (Tsuchinshan-ATLAS) and to infer possible differences. For this purpose, we rely on the method of telescopic wide-field polarimetric imaging of the comet in the visible to near-infrared domain in order to obtain the dependence of the degree of linear polarization (DoLP) of the coma and tail on the phase angle across a broad range. An off-the-shelf industrial grade polarization camera was used in combination with a telescope of short aperture ratio. These observations are complemented by T-matrix and Discrete Dipole Approximation modeling using the MSTM5 and DDSCAT software framework, respectively, for simulation of light scattering by dust particles of fractal agglomerate and agglomerate debris morphology. Our observations indicate that the coma exhibits a high maximum DoLP of $0.34$, which is further exceeded by a factor of about two by the DoLP of the comet's tail. Our modeling results suggest a 50:50 olivine-carbon composition. The fraction of agglomerate debris was found to be $50$\% in the coma and possibly higher in the tail. The differences between coma and tail in the observed maximum DoLP and the phase angle at which it occurs can be explained by a predominance of particles with radii larger than $0.6~\mu$m in the coma vs.\ smaller sub-micrometer particles close to the Rayleigh limit in the tail, assuming power-law size distributions with exponents of $2$ and $5$, respectively. Our results are consistent with smaller particles being transported from the coma into the tail more efficiently than larger particles by the solar radiation pressure. The possibly larger agglomerate debris fraction in the tail than in the coma may be a dynamical effect due to the mass difference between porous and compact particles of similar size.}
\section{Introduction}
\label{Introduction}
Ground-based telescopic polarimetry of comets is a widely used tool to assess the properties of their dust and gas hulls. Commonly, such studies involve measuring the degree and position angle of linear polarisation (DoLP and AoLP) of the coma. \citet{Kiselev2015} provided a profound introduction by covering the measurement techniques of aperture polarimetry, imaging polarimetry, and spectropolarimetry, and by discussing the different polarimetric classes of cometary comae essentially defined by the maximum DoLP value $P_{\rm max}$ they exhibit at phase angles near $90^\circ$, as well as by the wavelength dependence of the DoLP. They pointed out that due to the low number density of dust particles in cometary comae, multiple scattering between particles can be neglected so that their polarization behavior is governed by the single-scattering properties of the particles. High DoLP values were explained by a predominance of very small particles and/or highly absorbing carbon-rich particles. They described the model of \citet{Kolokolova2010} of a mixture of aggregated and solid particles to reproduce observational data by numerical modeling. In particular, they suggested physical explanations for the view that cometary dust particles are aggregates of hundreds or thousands of monomers with individual sizes smaller than a micrometer. \citet{Kiselev2017} compiled a database of polarimetric observations of a large number of comets from the literature and also from previously unpublished sources..

\citet{Levasseur2018} provided an extensive overview of the current knowledge about the chemical composition, mineralogy and physical properties of cometary dust, based on photometric, spectral and polarimetric telescopic observations as well as results from the Stardust and Rosetta missions. Regarding the structural properties of the dust particles, their porosities cover a broad range between 50\% and more than 90\%, where both compact and aggregate particles are of irregular shape. Spacecraft measurements provided particle sizes ranging from about $1~\mu$m to more than $1$~mm, where especially the highly porous particles were found to be fractal agglomerates composed of smaller constituents of sub-micrometer size. The review by \citet{Kolokolova2024} described photometric, spectral and polarimetric observation results obtained for different types of comets, numerical techniques for modelling the brightness and polarization of light scattered by cometary dust, and the effect of electromagnetic radiation on the dynamical behavior of dust particles. As important methods, they mentioned the multiple sphere T-matrix (MSTM) method for modeling fractal agglomerate (FA) particles and the discrete dipole scattering (DDSCAT) method for modeling agglomerate debris (AD) particles. According to this classification, the FA type of particles consists of multidisperse spheres with a given size distribution that stick to each other, thus forming a complex-shaped agglomerate that can be described by a fractal dimension. An AD particle is constructed by carving out parts from a spherical shape to generate a complex-shaped particle. \citet{Kolokolova2024} also provided a summary of still more realistic complex particle structures and their mixtures \citep[e.g.,][]{Halder2021}, where ambiguities induced by the large number of model parameters may arise, though, when attempting to fit the model to observational data. As a general result, it was considered by \citet{Kolokolova2024} that an explanation of ground-based photometric and polarimetric observations is likely to be achieved when assuming that the building blocks (monomers) constituting the dust particles are of submicrometer size.

An example of high-accuracy polarimetric cometary coma observations can be found in the study by \citet{Kwon2018} of the exceptionally short-periodic comet 2P/Encke. They measured the continuum DoLP of the dust along with the DoLP values of the gases C$_2$, NH$_2$ and CN. They found a high maximum DoLP of about $0.4$ for the dust in the coma, which they explained by the presence of relatively large and compact low-albedo dust particles with weak secondary scatttering. This particle morphology was suggested to be formed by removal of small dust particles from the coma by gas drag and by the compaction of porous agglomerates into more solid particles by the solar heat induced sublimation of volatiles. Near-infrared polarimetric observations of the near-Earth object 252P/LINEAR in the infrared J, H and K$_{\rm S}$ bands were conducted by \citet{Kwon2019}. They detected an abrupt increase of the DoLP accompanied by a bluing of the photometric and polarimetric colour in the course of an activation event. That behavior was explained by a scarcity of small porous particles and a predominance of large compact dust particles (governed by geometric optics) in the coma, ascribed to intense solar heating of the comet nucleus while located close to the Sun. \citet{Kwon2019} concluded that weathering or ``evolution'' of dust particles from porous to compact morphology might occur over time on the surface of the nucleus.

\citet{Bagnulo2024} stated that short-period comets that remain close to the Sun typically show low $P_{\rm max}$ values \citep{Kolokolova2007}. From this finding it is inferred that the comae of such comets are predominantly composed of highly compact dust particles. The synoptic analysis by \citet{Kwon2021} of a large set of comets with respect to their polarization behavior in the R (red) and K (near-infrared) bands and their thermal infrared silicate emission suggested that the polarization properties in the visible domain are mainly governed by the properties of the monomers constituting the dust aggregates, whereas the near-infrared polarization mostly depends on their porosity, with the dust composition being of minor relevance.

In several observational studies, spatially resolved DoLP maps of cometary comae were constructed. \citet{Ivanova2017} showed measurements of the wavelength-dependent DoLP of the coma of the comet C/2009~P1 (Garradd) in the visible and near-infrared domain. The DoLP was found to be nearly wavelength-independent beyond 550~nm and to decrease by about 50\% from the coma center to a distance of 40,000~km. Spectral, photometric and polarimetric observations of the coma of the distant ($>4~$AU from the Sun) comet C/2011~KP36 (Spacewatch) conducted by \citet{Ivanova2021} in the negative polarization branch indicated an increase of the absolute DoLP value by a factor of five between locations near the coma center and at distances of 20,000--50,000 km. Numerical modeling indicated a predominance of particles in the size range of several micrometers.

While a variety of detailed polarimetric studies were carried out on cometary comae, only a limited number of them were focused on cometary tails. The early visual work by \citet{Wright1881} revealed that the apparent length of the tail of comet B.1881 showed a strong dependence on the polarizer orientation due to a high DoLP of the tail. \citet{Matyagin1968} measured the DoLP and AoLP of the bright comet C/1965~S1 (Ikeya-Seki) by performing densitometry on photographs, obtaining DoLP values of up to 0.5--0.8 in the tail. \citet{Weinberg1976a} performed photometric scans of the tail of C/1965~S1 and the sky around it along profiles parallel to the horizon using eight different narrow-band filters in order to measure the DoLP and the AoLP, where phase angle dependent DoLP values between $-0.42$ and $+0.22$ were found. However, their observations were conducted at low elevations above the horizon between only $2^\circ$ and $20^\circ$, implying potential difficulties in correcting for elevation-dependent absorption by the Earth's atmosphere. Attempts were made to model the polarization behavior of the dust in the tail using Mie scattering, assuming a power-law size distribution of spherical particles. With the same observational setting, \citet{Weinberg1976b} examined the reversal of polarization in the tail of comet C/1965~S1 by determining the location of the point where the DoLP is zero and the AoLP exhibits a discontinuous change by $90^\circ$. More recently, \citet{Kwon2017} conducted photometric and polarimetric observations of the non-periodic comet C/2013~US10 in the visible and near-infrared domain, where the DoLP of the dust tail was found to be similar to that of the coma. For the coma of the periodic comet 67P/Churyumov-Gerasimenko, a sharp decrease of the DoLP at distances from the nucleus between 1500 and 5000~km followed by a gradual increase reaching a maximum at 40,000~km was observed by \citet{Rosenbush2017} in the positive and negative polarization branch. They explained this behavior by an increasingly steep size distribution of the dust particles at increasing distance from the nucleus. Furthermore, their data showed that the DoLP of the brightest parts of the tail is similar to that of the outer coma. Measurements by \citet{Ivanova2019} of the DoLP in the negative polarization branch of the coma and tail of the distant comet C/2014~A4 (SONEAR) revealed an increase of the absolute DoLP value by a factor of four between the coma center and the tail at 200,000~km distance in antisolar direction. Based on numerical modeling assuming rough spheroidal particles, they concluded that the dust particles in the coma have sizes in the sub-micrometer and micrometer range with variations of polarization being largely due to compositional differences. \citet{Bagnulo2021} presented DoLP profiles of the coma and tail of the interstellar comet 2I/Borisov in the visible domain measured with increasing apertures centered on the nucleus, indicating that the absolute DoLP values increase with increasing aperture both in the negative and the positive polarization branch. In the study by \citet{Nezic2022}, data of the COR2 coronographs on-board the spacecraft STEREO-A and STEREO-B were used to analyze the DoLP along the tail of the Kreutz comet C/2010~E6 (STEREO). At phase angles around $110^\circ$ they found DoLP values of up to about $0.3$ in the tail, where the DoLP increased monotonically with increasing distance from the nucleus.

In this paper we examine the bright comet C/2023~A3 (Tsuchinshan-ATLAS), which is ``dynamically new'' \citep{Cambianica2025} as it visited the inner solar system for the first time during its visibility period in 2024. Inferring the structural properties of the dust particles in the coma and tail of this comet from the polarization behavior of the scattered light will thus provide insights into the primordial matter that formed the planetary bodies of the solar system. Hence, we describe wide-field imaging polarimetric observations of the coma and in particular the tail of C/2023~A3 across a broad range of phase angles. Using the T-matrix method and the Discrete Dipole Approximation technique, we attempt to explain our observational data by numerical simulations of the polarization behavior of the dust agglomerates in the coma and tail in order to assess their respective structural and compositional properties. In this context, also the abovementioned dependence of the DoLP on the particle size distribution \citep[e.g.,][]{Rosenbush2017} will be examined.
\section{Data acquisition and analysis methods}
\label{DataAcquisitionAnalysisMethods}
\subsection{Data acquisition}
\label{sec:DataAcquisition}
For acquisition of the polarization imaging data of comet C/2023~A3 examined in this study, a monochrome industrial camera DZK~33UX250 without cooling was used, manufactured by the company The Imaging Source. To the individual pixels of the camera sensor, polarization filters rotated by an angle $\theta$ of $0^\circ$, $45^\circ$, $90^\circ$ and $135^\circ$, respectively, relative to the horizontal image direction are attached \citep{TIS2019}. Thus, from each $2\times 2$ subset of pixels the unpolarized intensity, the DoLP and the AoLP relative to the direction of the pixel rows can be derived. For all observations but one, the camera was mounted on a Newton reflector telescope of $150$~mm diameter with a focal length of $600$~mm. The resulting intensity, DoLP and AoLP images have a scale of $2.4$~arcseconds/pixel with a total field of view of $48 \times 40$ arcminutes. The location of the telescope is in Wetter (Ruhr) near Dortmund, Germany. The observation of Nov~14 was conducted with a small $30/135$~mm refractor at $10.5$~arcsec/pixel at the ARIES Observatory on Manora Peak near Nainital, India. With this setup we acquired broadband data in order to maximize the signal-to-noise ratio, where the sensitivity of the camera sensor is highest around 600~nm \citep[Fig.~\ref{fig:SpectrumPostPerihelion_Sensitivity},][]{TIS2019}. Additional observations of the coma DoLP were conducted on Dec~05, 2024, and on Mar~06, 2025, with the 1.2~m reflector at Mount Abu Observatory, Rajasthan, India, which was equipped with the EMPOL imaging polarimeter \citep{Ganesh2020} upgraded with a new, electrically cooled CCD camera. The acquisition details of our polarimetric images are given in Tables~\ref{tab:Observations1} and~\ref{tab:Observations2}.
\begin{table*}[tbp]
\caption{Wide-field broadband imaging polarimetric observations conducted with industrial polarization camera. Phase angles and distances (in AU) from the JPL Horizons Ephemeris \citep{Giorgini1996}.}
\label{tab:Observations1}
\centering
\begin{tabular}{lcrccrrrr}
Date & UT & Integration & Phase & Binning & DoLP & DoLP & Earth & Solar\\
(2024) & (start) & time [min] & angle & for tail & coma & tail & dist. & dist.\\
\hline
Oct~23 & 19:02 & 70 & $85.5^\circ$ & -- & $0.298$ & $0.667$ & 0.70 & 0.77 \\
Oct~25 & 19:45 & 59 & $78.6^\circ$ & -- & $0.316$ & $0.495$ & 0.77 & 0.81 \\
Oct~26 & 18:43 & 98 & $75.6^\circ$ & -- & $0.307$ & $0.583$ & 0.80 & 0.82 \\
Oct~27 & 18:23 & 20 & $72.3^\circ$ & -- & $0.246$ & $0.782$ & 0.83 & 0.84 \\
Oct~31 & 18:39 & 44 & $63.3^\circ$ & -- & $0.225$ & $0.496$ & 0.97 & 0.92 \\
Nov~03 & 17:31 & 98 & $57.6^\circ$ & -- & $0.214$ & $0.523$ & 1.07 & 0.98 \\
Nov~04 & 18:35 & 63 & $55.9^\circ$ & -- & $0.151$ & $0.336$ & 1.11 & 1.00 \\
Nov~05 & 17:46 & 59 & $54.3^\circ$ & -- & $0.150$ & $0.336$ & 1.14 & 1.02 \\
Nov~14 & 13:35 & 74 & $42.7^\circ$ & -- & $0.077$ & $0.366$ & 1.44 & 1.18 \\
Nov~29 & 19:12 & 17 & $30.1^\circ$ & $2\times$ & $0.046$ & $0.093$ & 1.92 & 1.45 \\
Nov~30 & 18:22 & 56 & $29.4^\circ$ & $2\times$ & $0.074$ & $0.094$ & 1.95 & 1.47 \\
Dec~01 & 16:43 & 5 & $28.8^\circ$ & $2\times$ & $0.062$ & $0.127$ & 1.98 & 1.49 \\
Dec~07 & 18:40 & 9 & $25.3^\circ$ & $4\times$ & $0.036$ & $0.145$ & 2.15 & 1.59 \\
Dec~08 & 17:36 & 22 & $24.8^\circ$ & $4\times$ & $0.021$ & $0.229$ & 2.18 & 1.60 \\
Dec~27 & 18:02 & 27 & $16.7^\circ$ & $2\times$ & $-0.034$ & $-0.094$ & 2.65 & 1.91 \\
Dec~28 & 17:27 & 50 & $16.4^\circ$ & $2\times$ & $-0.026$ & $-0.114$ & 2.67 & 1.93 \\
\end{tabular}
\end{table*}
\begin{table*}[tbp]
\caption{Imaging polarimetric observations conducted with the Mount Abu 1.2~m reflector and upgraded EMPOL instrument. Earth and solar distance in AU. Phase angles and distances from the JPL Horizons Ephemeris \citep{Giorgini1996}. Filter band data from \citet{Fukugita1996}.}
\label{tab:Observations2}
\centering
\begin{tabular}{lccccccc}
Date & UT & Filter & Phase & Binning & DoLP & Earth & Solar \\
 & (start) & (SLOAN) & angle & & coma & dist. & dist. \\
\hline
Dec~05, 2024 & 13:20 & i ($770 \pm 75$~nm) & $26.5^\circ$ & $4\times$ & $0.0141 \pm 0.0064$ & 2.09 & 1.55 \\
 & & z ($910 \pm 60$~nm) & & & $0.0356 \pm 0.0108$ & & \\
Mar~06, 2025 & 00:02 & r ($625 \pm 70$~nm) & $14.1^\circ$ & $4\times$ & $< -0.0073$ & 2.89 & 3.50 \\
\end{tabular}
\end{table*}

\begin{figure*}[tbp]
\centering
\includegraphics[width=0.9\textwidth]{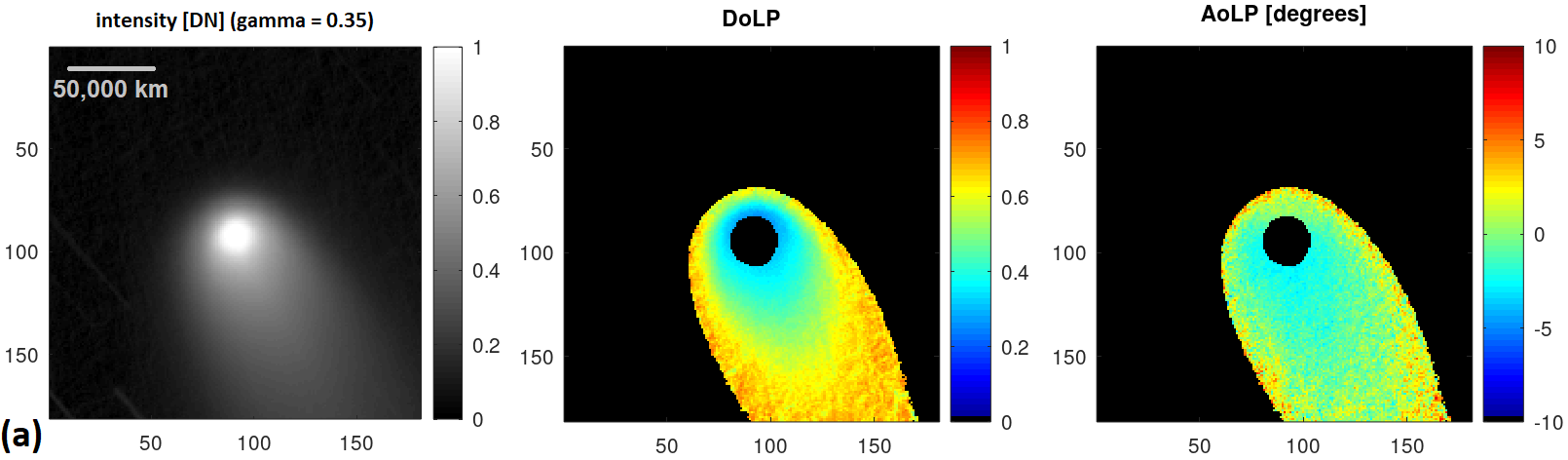}\\
\includegraphics[width=0.375\textwidth]{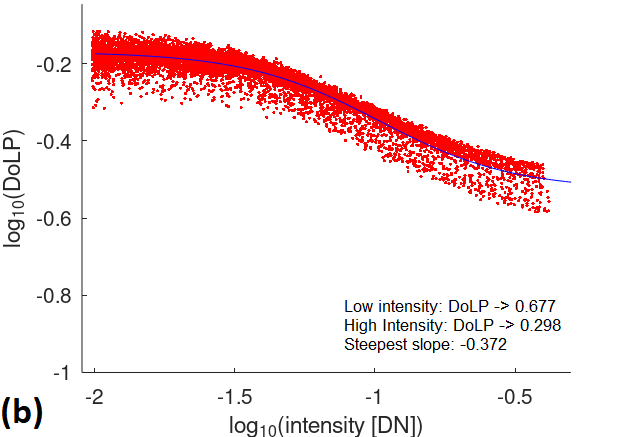}\includegraphics[width=0.28\textwidth]{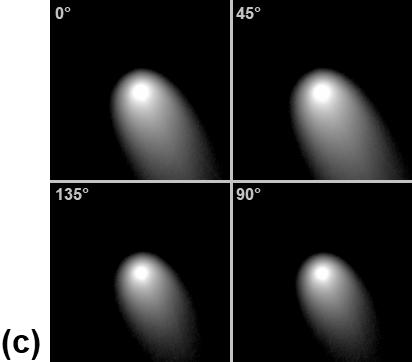}
\caption{(a)~Pixel intensity $F$ (in DN, clipped at~1 due to sensor saturation, contrast-enhanced with gamma value of $0.35$), DoLP $P$ and AoLP $W$ (in degrees) of comet C/2023~A3 on Oct~23. Image axes denote pixel coordinates, with a scale of 2.4~arcsec or 1200~km per pixel. (b)~Measured $\lg(P)$ vs.\ $\lg(F)$ values (red dots) and fitted hyperbolic tangent function (blue curve). (c)~Oct~23 image quadruple of C/2023~A3 at polarizer orientations of $0^\circ$, $45^\circ$, $90^\circ$ and $135^\circ$. The pixel intensities are scaled logarithmically for better visibility of the tail.}
\label{fig:FPW_Oct23}
\end{figure*}

During each observation run with the polarization camera, we acquired several tens to hundreds of individual video frames taken with the maximum possible exposure time of $4$~s. From each video frame the sub-frames corresponding to the four polarization directions were extracted. The sub-frames of each polarization direction were coregistered and stacked with the software Autostakkert3 \citep{Kraaikamp2023}, resulting in one image of $16$~bits pixel depth per polarization direction. Subsequently, the $45^\circ$, $90^\circ$ and $135^\circ$ images were then coregistered at subpixel accuracy with respect to the $0^\circ$ image, where the image transformations (translations) were computed by minimizing the sum of squared differences between the high-frequency components of each image pair \citep{Bhatt2023}. To the four intensity values $I(\theta)$ at each pixel position of the resulting quadruple of coregistered images, a sinusoidal function of the form
\begin{equation}
\label{eq:Itheta}
I(\theta)=F+G\cos\left[2\left(\theta-W\right)\right]
\end{equation}
was fitted, yielding images of the intensity $F$, the DoLP $P=|G|/F$, the AoLP $W$, and the residual root-mean-squared error (RMSE) $E$ of the fit. As an example, the results of the October~23 observation are shown in Fig.~\ref{fig:FPW_Oct23}. For this observation and all other ones for which the coma was saturated in the image data due to its excessive brightness, a hyperbolic tangent function was fitted to the $\lg(P)$ vs.\ $\lg(F)$ measurements in order to obtain a robust extrapolation of the DoLP to very low and very high intensities, respectively (Fig.~\ref{fig:FPW_Oct23}b). The hyperbolic tangent function was chosen for its smooth asymptotic behavior which prevents unrealistic DoLP values at very low or high intensities. Fig.~\ref{fig:FPW_Oct23}c displays the four subframes corresponding to the polarizer orientations of $0^\circ$, $45^\circ$, $90^\circ$ and $135^\circ$, where the varying length of the visible tail indicates a high DoLP.

To avoid using data near saturation, the DoLP was only computed for pixels with 16-bit intensities below $35,000$, corresponding to $F_{\rm max}\approx 0.53$ when the pixel intensity is normalized to the interval $[0,1]$. Low-SNR data were excluded by neglecting pixels with $F/E<10$ (see Sect.~\ref{sec:ErrorAnalysis}), corresponding to $F_{\rm min}\approx 0.01$ with pixel intensity normalized to $[0,1]$. Later observations at the end of November and during December 2024 showed a strong decrease in apparent size of the comet, whose areal brightness also decreased due to its increasing solar distance. Due to the resulting small number of pixels covered by the comet, it was not always possible to fit the hyperbolic tangent function to the $\lg(P)$ vs.\ $\lg(F)$ measurements. In these cases, the DoLP of the tail was measured by averaging over all pixels with $F\in\left[F_{\rm min}, 1.58\times F_{\rm min}\right]$, an interval covering $0.2$ orders of magnitude in decadal logarithmic scale. Similarly, the average DoLP over all pixels with $F\in\left[F_{\rm max}/1.58, F_{\rm max}\right]$ was considered as the DoLP of the coma. In order to still be able to measure the DoLP of the tail, binning of the images by a factor of $2$ or $4$ was performed (Table~\ref{tab:Observations1}) to increase the SNR.

\begin{figure}[tbp]
\centering
\includegraphics[width=0.7\textwidth]{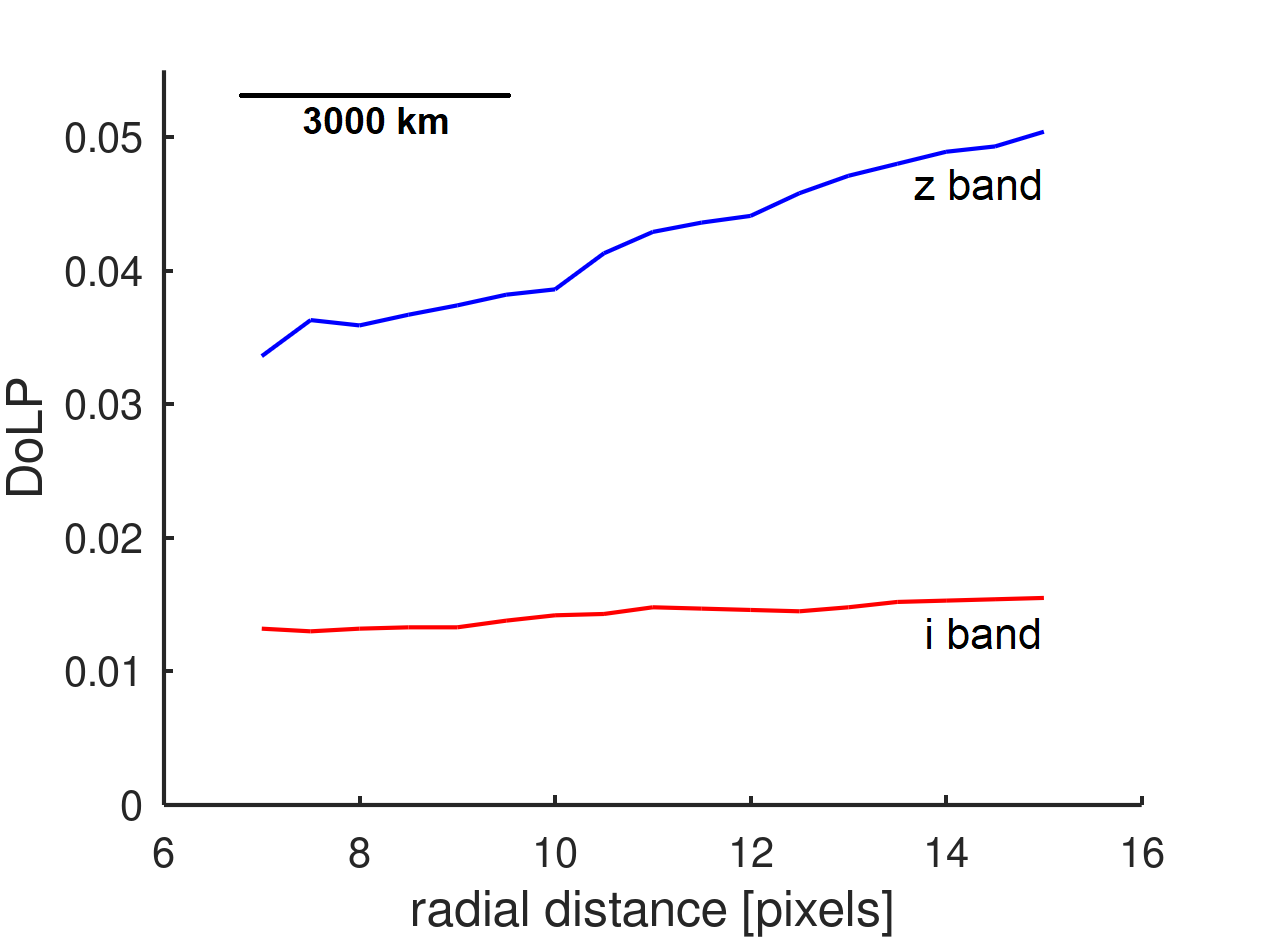}
\caption{Dependence of DoLP on radial distance from the nucleus, measured on Dec~05 with Mount Abu 1.2~m reflector and EMPOL. Image scale is 0.72~arcsec or 1100~km per pixel.}
\label{fig:RadialDoLP_Dec05}
\end{figure}

Fig.~\ref{fig:FPW_Oct23}a demonstrates that the DoLP is larger in the tail and the outer parts of the coma than in the inner coma close to the nucleus. Our observations with the Mount Abu 1.2~m reflector equipped with the EMPOL imaging polarimeter, conducted on Dec~05 in the SLOAN~i and~z bands centered at $763$ and $913$~nm, respectively, confirm this behavior (Fig.~\ref{fig:RadialDoLP_Dec05}). The radial distance from the nucleus, though, is restricted to about 11~arcseconds due to the comparatively narrow field of view. 
\subsection{Error analysis}
\label{sec:ErrorAnalysis}
This section explains how we estimate the statistical error of the inferred DoLP $P$ in dependence on the signal-to-noise ratio $\mathrm{SNR}_F$ of the intensity $F$ based on a Monte-Carlo model of the image acquisition process \citep[for a general description of the estimation of error quantities with Monte Carlo methods see, e.g.,][]{Harding2014}. For illustration, Fig.~\ref{fig:FPW_Oct23}c depicts a quadruple of frames of the coma and tail of C/2023~A3 for polarizer orientations $\theta$ of $0^\circ$, $45^\circ$, $90^\circ$ and $135^\circ$. The frames are pixel-synchronous, so that a least-squares fit of Eq.~\ref{eq:Itheta} to the four pixel intensities at a specific location directly yields the intensity $F$, DoLP $P$, AoLP $W$, and residual root-mean-squared fitting error $E$ at that same location. For our error analysis, we construct a set of four artificial frames of a uniformly lit area for the abovementioned $\theta$ values. Each artificial frame has $100 \times 100$ pixels with a $\theta$-dependent uniform grayvalue $I(\theta)$ according to Eq.~\ref{eq:Itheta} with given values for $F$, $G$ and $W$, where Gaussian noise with a standard deviation of $2\sigma$ is added to the grayvalues of each frame. Fitting Eq.~\ref{eq:Itheta} to each pixel location of this set of frames yields images of $F$, $P$, $W$ and $F/E$, which allow for computing the mean values and uncertainties (standard deviations) of these quantities in dependence on the noise level $\mathrm{SNR}_F$ of the intensity $F$. As the intensity $F$ fitted using Eq.~\ref{eq:Itheta} equals the average of the four corresponding grayvalues, and the noise of the average of $M$ noisy measurements decreases proportionally to $M^{-1/2}$ with $M=4$, one obtains $\mathrm{SNR}_F=F/\sigma$. It is not straightforward to measure the noise level $\sigma$ of our $F$ images of the comet directly, and the effect of photon statistics even leads to a dependence of $\sigma$ on $F$. However, our simulations for $\mathrm{SNR}_F$ values between 10 and 1000 indicate that the ratio $F/E$ is a proxy for the signal-to-noise ratio with $\mathrm{SNR}_F \approx 0.8 \times (F/E)$.

\begin{figure}[tbp]
\centering
\includegraphics[width=0.7\textwidth]{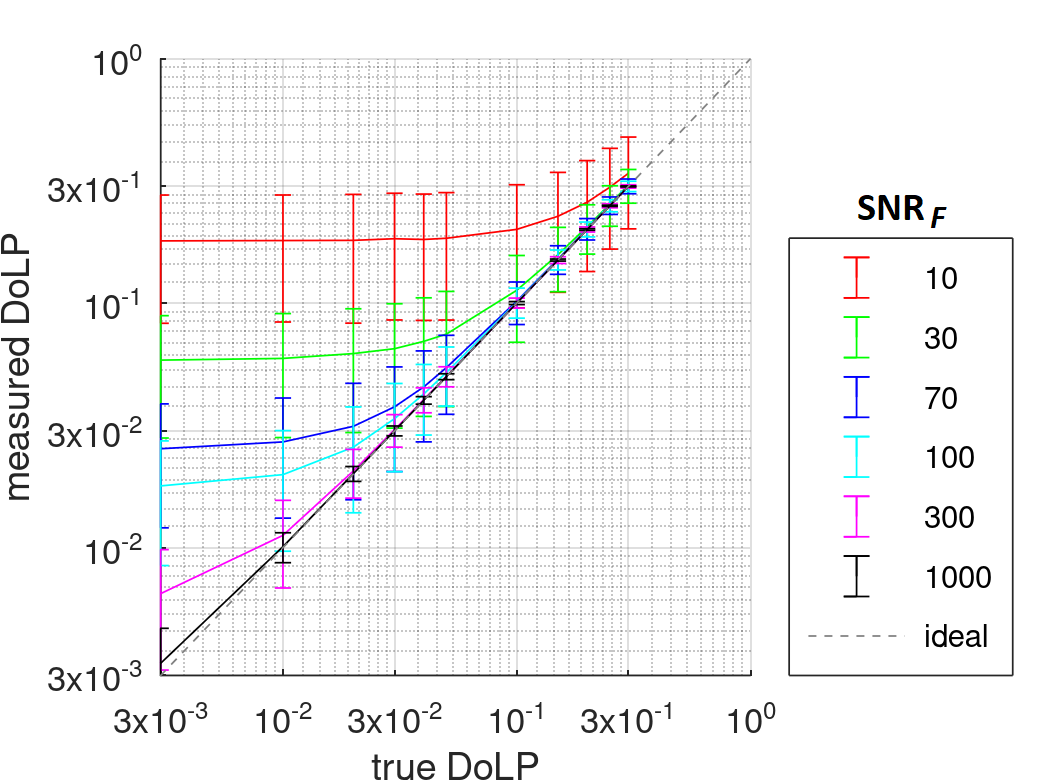}
\caption{Measured vs.\ true DoLP for different $\mathrm{SNR}_F$ values, as obtained by Monte-Carlo based error analysis.}
\label{fig:ErrorAnalysisResults}
\end{figure}

Even in the absence of linear polarisation, a fit of Eq.~\ref{eq:Itheta} to noisy grayvalues of the $\theta$-dependent frames leads to a non-zero best-fit value of $G$ and thus to a positive estimated DoLP $P$, whose value increases with decreasing $\mathrm{SNR}_F$. Hence, only DoLP values exceeding an $\mathrm{SNR}_F$-dependent threshold can be measured reliably. The simulated relation between the true and the measured DoLP is shown in Fig.~\ref{fig:ErrorAnalysisResults} for a range of typical $\mathrm{SNR}_F$ values between $10$ and $1000$. The diagram shows that the DoLP must exceed a $\mathrm{SNR}_F$-dependent ``detection threshold'' $\Theta$ to be measurable. If the true DoLP becomes smaller than $\Theta$, the measured DoLP is merely an upper limit to the true DoLP. For $\mathrm{SNR}_F$ values of $10$ and $100$, typical signal-to-noise ratios of the comet's tail and central coma, respectively, values of $\Theta$ around $0.2$ and $0.02$ are found. Besides this systematic effect, the statistical measurement error (one standard deviation) was found to correspond to approximately $0.1$ and $0.01$, respectively, for DoLP values larger than $\Theta$. More specifically, if for example the measured DoLP is $0.03$ and $\mathrm{SNR}_F<70$, then the true DoLP is below the corresponding threshold $\Theta$ so that it cannot be detected reliably. For $\mathrm{SNR}_F=70$ it is possible to determine the true DoLP with a statistical measurement error of $\pm 0.017$. For $\mathrm{SNR}_F=1000$, the systematic error becomes negligible for DoLP values exceeding $0.01$, and the statistical measurement error for a true DoLP of $0.03$ reduces to $\pm 0.0014$. Representative values of the DoLP detection threshold $\Theta$ and the statistical measurement error are listed in Table~\ref{tab:ErrorAnalysis}.
\begin{table}[tbp]
\caption{DoLP detection threshold and statistical measurement error (one standard deviation) for typical values of SNR$_F$.}
\label{tab:ErrorAnalysis}
\centering
\begin{tabular}{c|ll}
SNR$_F$ & DoLP detection & DoLP \\
 & threshold $\Theta$ & std.\ dev.\ \\
\hline
10 (20~dB) & 0.18 & 0.095--0.135 \\
30 (30~dB) & 0.06 & 0.030--0.045 \\
100 (40~dB) & 0.02 & 0.009--0.015 \\
300 (50~dB) & 0.006 & 0.003--0.005 \\
1000 (60~dB) & 0.002 & 0.001--0.002 \\
\end{tabular}
\end{table}
\subsection{Polarization modeling}
\label{sec:PolarizationModeling}
An important morphological class of cometary dust particles corresponds to the fractal aggregate (FA) type \citep[e.g.,][]{Kiselev2015}. FAs consisting of spherical monomers with a given log-normal distribution of their radii were constructed using the FracVAL algorithm, where the monomers form a chain structure whose shape is defined by a fractal dimension $D_f$ between 1 (linear chain) and 3 (densely packed cluster) \citep{Moran2019}. We used the Python implementation of FracVAL by \citet{Schauten2025}. Setting the number $N$ of monomers per agglomerate to 1, 2, 4, 8, 20, 48, 80, 128, 256, 384 and 512 with $D_f$ values of $1.4$, $1.6$, $1.8$, $2.0$ and $2.2$, respectively, resulted in a set of 40 different agglomerate shapes with $N \geq 8$ (Fig.~\ref{fig:Agglomerates}) and four agglomerate shapes with $N \leq 4$. In the latter case, ``agglomerates'' with $N=1$ and $N=2$ only have one possible configuration, respectively, while for $N=4$ we used the linear and the tetrahedral configuration. The constructed agglomerate shapes were scaled uniformly such that the mean monomer radius $r_{\rm mean}$ obtained values of 112.5~nm, 125~nm and 137.5~nm. Constant values of $1.05$ and $1.25$ were used for the logarithmic standard deviation of the monomer radii and for the fractal prefactor, respectively. The radius of gyration $R_g$ of the agglomerates decreases with increasing fractal dimension $D_f$ and spans the intervals of $1.0$--$74\times r_{\rm mean}$ for $D_f=1.4$ and $1.0$--$15\times r_{\rm mean}$ for $D_f=2.2$, respectively. A further simulation parameter is the lower cutoff threshold $c$ of the agglomerate size distribution, whose effect is that agglomerates with $R_g<c$ are assumed to be absent in the simulation. Notably, the fact that only relatively small particles were modeled due to computational constraints is a possible deficiency of our modeling (cf.\ Sect.~\ref{sec:DiscussionModeling} for further discussion).

We computed the polarization behavior of the agglomerates using the MSTM5 software framework \citep{Mackowski2022}, where we assumed that the agglomerates consist of pure olivine or pure amorphous carbon \citep{Zubko2016} and used the complex refractive indices $m_{\rm olivine}=1.85+0.00087i$ \citep{Fabian2001,ARIA2025} and $m_{\rm carbon}=1.99+0.222i$ \citep{Rouleau1991}. We assumed a particle size distribution obeying a power law, i.e., the number of agglomerates is proportional to $R_g^{-n}$ with $n$ as the power law exponent \citep[e.g.,][]{Zubko2016,Halder2021}. Thus, with increasing value of $n$ the relative influence of large dust particles on the observed polarization behavior becomes weaker. As the coma and tail of a comet can be treated as optically thin media \citep{Halder2021}, the DoLP phase curve of a specific mixture of olivine and carbon is given by a correspondingly weighted average of the DoLP phase curves obtained for the pure compositions. As our observations were conducted without filter, our simulations were performed for a wavelength of 600~nm that corresponds to the maximum sensitivity of the camera sensor \citep{TIS2019}.

Our MSTM-based simulations show that FA alone cannot describe the observed polarization of the coma and tail under physically reasonable constraints (Sect.~\ref{sec:ModelingResults}). Hence, in this study we have also included agglomerate debris (AD) particles \citep[e.g.,][]{Draine1994,Draine2013,Halder2021,Halder2022} as low-porous solids, in particular to compensate for the high $\alpha_{\rm max}$ value observed in the tail. We understand that the AD particles are a kind of controversy, as they do not represent the physics of particle growth. In this study, though, we hypothesize that the AD particles are not the fundamental building blocks of dust but rather the byproducts and/or remnants of large fractals that were highly processed by solar radiation. On the other hand, the traditional FA particles are also not free from controversy. In the FA model, we consider a cluster of spherical monomers that touch each other at a single contact point \citep{Moran2019}. In-situ measurements of dust particles reveal that the contact between the monomers is not a point, but the monomers tend to be of non-spherical shape and contact each other at multiple points or are fused to one another \citep[as shown, e.g., in Fig.~2 of][]{Mannel2019}. On the other hand, FAs as shown in Fig.~\ref{fig:Agglomerates} may not be strictly representative of cometary dust but rather correspond to basic building blocks of more complex, hierarchical structures \citep[][]{Kolokolova2018}. We furthermore understand that in cometary polarization, the monomer size of FAs plays a crucial role.

Thus, in this study we use FAs to represent high porosity particles and AD to represent low porosity particles. In addition to MSTM5, we thus used the REST software package \citep{Halder2022} to generate AD particles having a packing fraction of approximately 0.26 and computed the polarization properties using the Discrete Dipole Approximation \citep{Draine1994,Draine2013} for the size range of $0.1$--$2.0~\mu$m for 100 particle sizes and eight different particle realizations, using the complex refractive indices of olivine and amorphous carbon discussed above for a wavelength of 600~nm.
\section{Results}
\label{sec:Results}
\subsection{Observation results}
\label{sec:ObservationResults}
\begin{figure}[tbp]
\centering
\includegraphics[width=0.7\columnwidth]{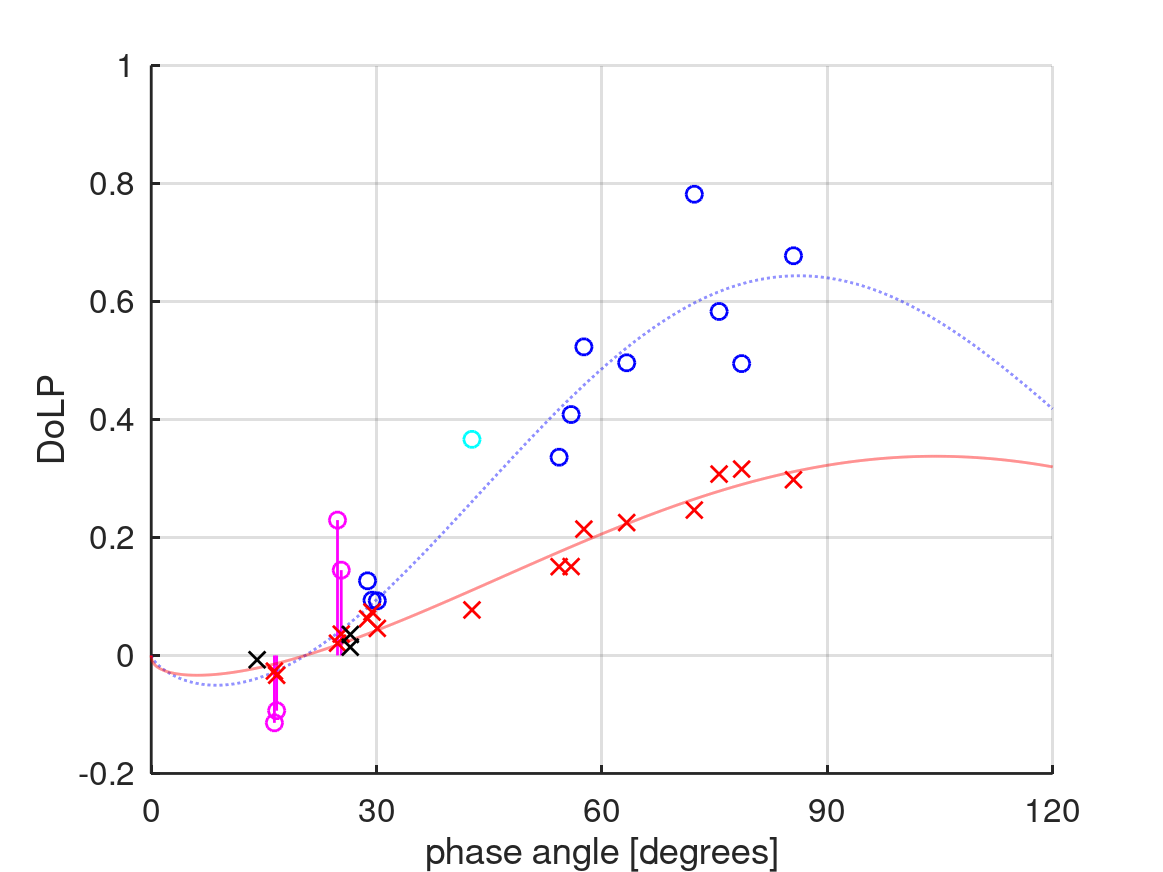}
\caption{Observed DoLP data of the coma (crosses) and tail (circles) of comet C/2023~A3, with empirical DoLP phase curves fitted as described in Sect.~\ref{sec:ObservationResults}. Red crosses indicate valid broadband DoLP measurements of the coma with the polarization camera, black crosses denote EMPOL measurements of Dec~05, 2024, in the SLOAN~i and~z band and of Mar~06, 2025, in the SLOAN r~band, respectively. Dark blue circles indicate valid broadband DoLP measurements of the tail, the cyan circle corresponds to an inaccurate measurement, and the magenta circles denote upper bounds of the tail's DoLP due to low $\mathrm{SNR}_F$ (Sect.~\ref{sec:ErrorAnalysis}). The blue phase curve was fitted to the measurements marked by blue circles only.}
\label{fig:DoLP_observed}
\end{figure}

The measured DoLP $P$ vs.\ phase angle $\alpha$ of the coma and the tail are shown in Fig.~\ref{fig:DoLP_observed} as crosses and circles, respectively. The tail measurement plotted in cyan color has a high uncertainty, and the measurements plotted in magenta color are upper limits to the respective DoLP values as a consequence of the systematic $\mathrm{SNR}_F$-dependent effect explained in Section~\ref{sec:ErrorAnalysis}. To the coma measurements, we fitted the empirical function
\begin{equation}
\label{eq:TrigFunction}
P(\alpha)=A\sin^B(\alpha)\cos^C(\alpha/2)\sin(\alpha-\alpha_{\rm inv})
\end{equation}
introduced by \citet{Lumme1993} and \cite{GoidetDevel1995}, performing a least-squares estimate of the parameters $A$, $B$, $C$ and $\alpha_{\rm inv}$ with all four parameters being constrained to non-negative values. The phase angle of inversion $\alpha_{\rm inv}$ is the phase angle at which $P=0$ and the plane of polarization rotates by an angle of $90^\circ$, effectively resulting in a change of the sign of $P$ from positive to negative. As we could acquire reliable tail measurements of $P$ only down to a phase angle of $28.8^\circ$, we used the same value of $\alpha_{\rm inv}$ for the tail measurements and only estimated the parameters $A$, $B$ and $C$. Furthermore, we derived from the fitted polarization phase curves the maximum DoLP $P_{\rm max}$, the phase angle $\alpha_{\rm max}$ with $P(\alpha_{\rm max})=P_{\rm max}$, and the slope $h=dP/d\alpha$ at $\alpha=\alpha_{\rm inv}$. These parameters are commonly used to describe the behavior of the positive polarization branch of comets, asteroids or the Moon \citep[e.g.,][]{Bagnulo2024}. The parameter values obtained for the coma and the tail, respectively, are listed in Table~\ref{tab:PolarizationParameters}. Notably, the estimated value of $\alpha_{\rm max}$ of the coma may be inaccurate by about $10^\circ$--$20^\circ$ because no DoLP measurements are available for phase angles exceeding $85.5^\circ$ that would have been able to better constrain the empirical function (Eq.~\ref{eq:TrigFunction}) around the maximum \citep[such measurements have been provided for the coma by][though]{Lim2025}. Furthermore, the limited coverage of our observations for small phase angles does not allow for a reliable estimation of the minimum DoLP $P_{\rm min}$ and the phase angle $\alpha_{\rm min}$ at which it occurs.
\begin{figure}[tbp]
\centering
\includegraphics[width=0.7\columnwidth]{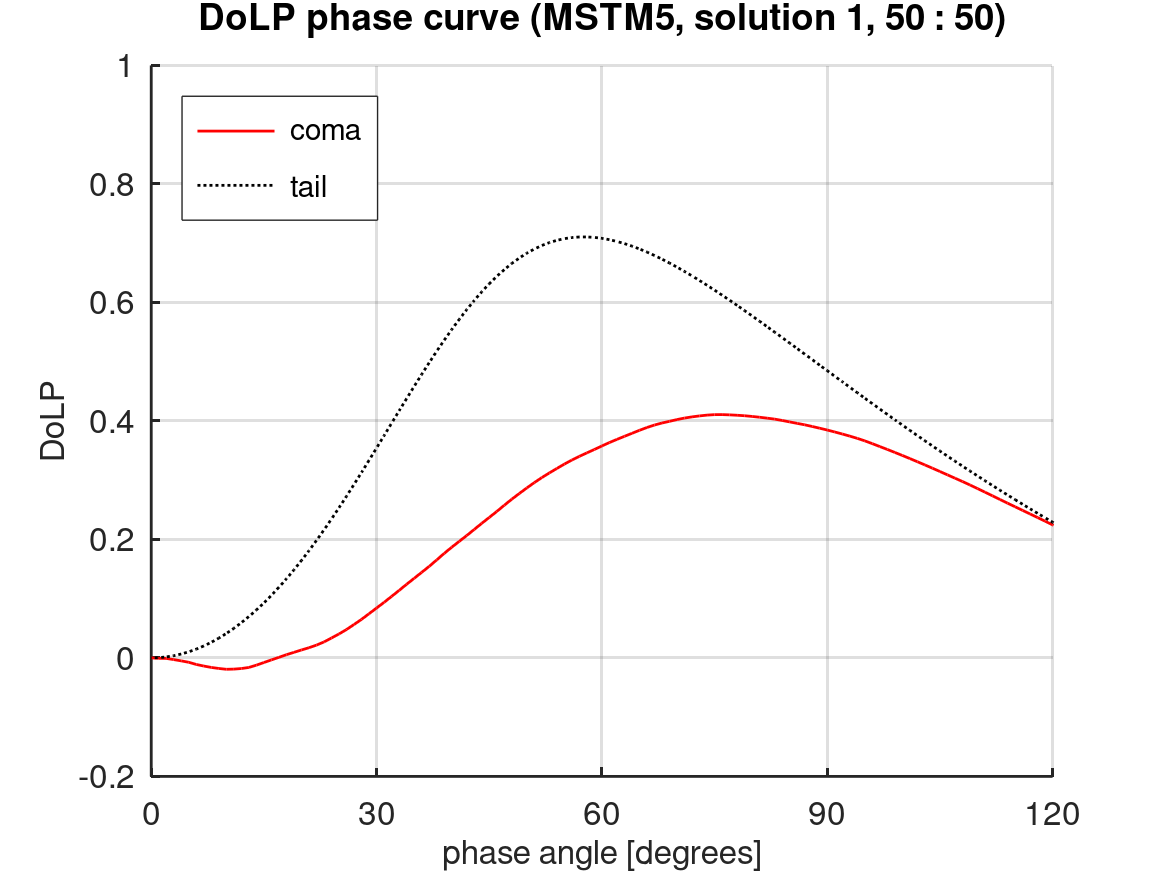}
\caption{MSTM5-based solution~1: different agglomerate size distributions for coma and tail, identical composition (50:50 olivine-carbon) and identical values of $r_{\rm mean}=137.5$~nm. See Sect.~\ref{sec:ModelingResults} for details.}
\label{fig:DoLP_modeled_sol1}
\end{figure}
\begin{figure}[tbp]
\centering
\includegraphics[width=0.7\columnwidth]{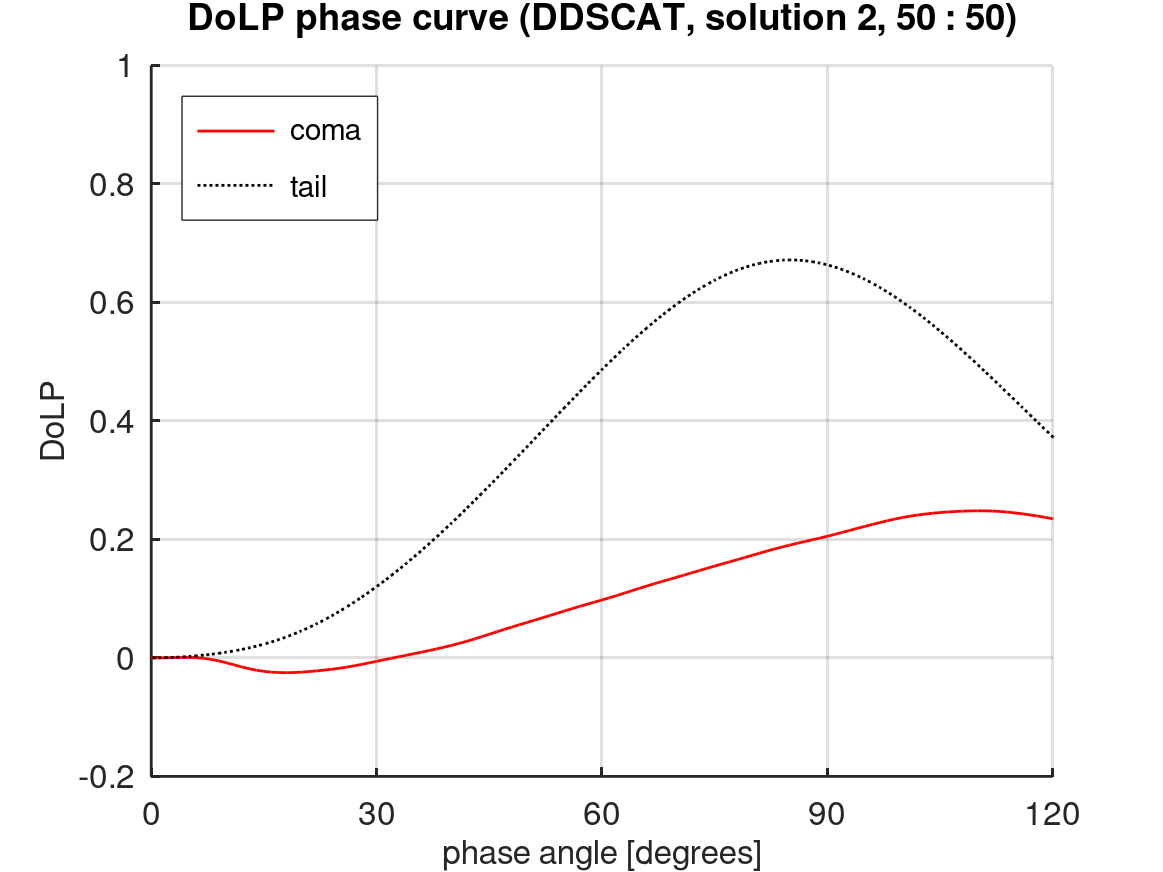}
\caption{DDSCAT-based solution~2: identical composition (50:50 olivine-carbon) and different size distributions for coma and tail. See Sect.~\ref{sec:ModelingResults} for details.}
\label{fig:DoLP_modeled_sol2}
\end{figure}
\begin{figure}[tbp]
\centering
\includegraphics[width=0.7\columnwidth]{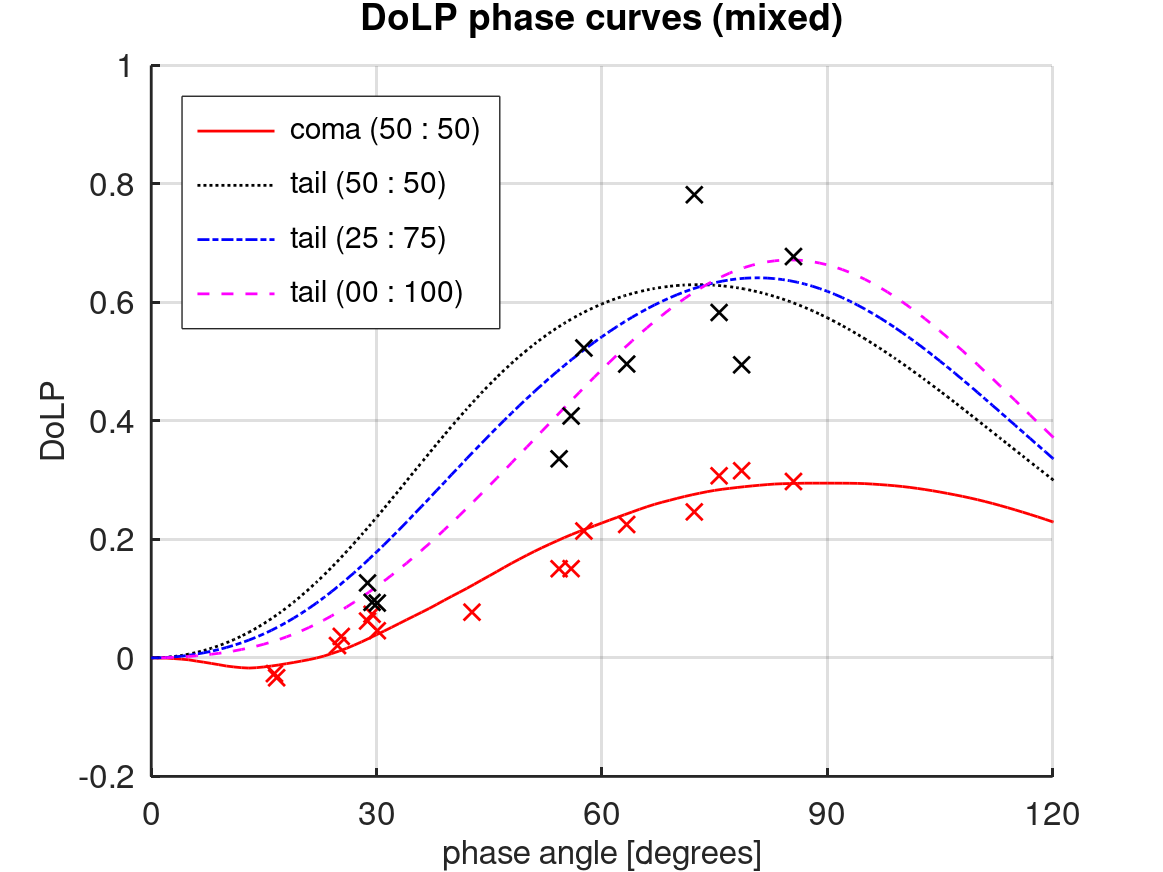}
\caption{Mixed solution~3: mixture between solution~1 (FA) and solution~2 (AD), with mixture coefficients of 50:50 in the coma and 50:50, 25:75 and 0:100 in the tail. Crosses denote our DoLP observations of the coma (red) and tail (black). See Sect.~\ref{sec:ModelingResults} for details.}
\label{fig:DoLP_modeled_sol3}
\end{figure}
\begin{table}[tbp]
\caption{Estimated polarization parameters of the coma and tail of comet C/2023~A3.}
\label{tab:PolarizationParameters}
\centering
\begin{tabular}{l|rrrr}
 & $P_{\rm max}$ & $\alpha_{\rm max}$ [deg] & $h$ [deg$^{-1}$] & $\alpha_{\rm inv}$ [deg]\\
\hline
coma & 0.34 & 104.4 & 0.0039 & 20.5\\
tail & 0.64 & 86.2 & 0.0079 & 20.5\\
\end{tabular}
\end{table}
\subsection{Modeling results}
\label{sec:ModelingResults}
For dust particles of FA morphology, the effects of the mean monomer radius $r_{\rm mean}$, the fractal dimension $D_f$ and the power law exponent $n$ of the particle size distribution on the DoLP phase curve $P(\alpha)$ are shown in Figs.~\ref{fig:ModelParameters1} and~\ref{fig:ModelParameters2}. For particles consisting of $N=1$ or $N=2$ monomers, the value of $D_f$ is meaningless, and always the same shapes were used independent of $D_f$. For $N=4$, the linear configuration was used for $D_f=1.4$ and the tetrahedral configuration otherwise. For olivine, an increase of $D_f$ from $1.4$ to $2.2$ leads to a reduction of $P_{\rm max}$ of at most $15\%$ and decreases with increasing $r_{\rm mean}$ and $n$. For carbon, $P_{\rm max}$ shows no clear dependence on $D_f$. Increasing the power law exponent $n$ from $2$ to $5$, i.e., increasing the fraction of small dust particles, leads to an increase of $P_{\rm max}$ of up to $20\%$ and a nearly $D_f$-independent behavior of $P(\alpha)$. Consequently, the model parameter that has the strongest influence on $P_{\rm max}$ is the mean monomer radius $r_{\rm mean}$. This is in accordance with the finding of \citet{Kwon2021} that the polarization of cometary dust in the visible (red) domain is mainly governed by the monomer properties.

The DoLP phase curves in Fig.~\ref{fig:DoLP_observed} show that the $\alpha_{\rm max}$ value of the coma is larger than that of the tail, which remains true in the light of the $\alpha_{\rm max}$ value of $90^\circ$--$95^\circ$ estimated for the coma by \citet{Lim2025}. In our simulation analyses we found that the observed differences in $P_{\rm max}$ and $\alpha_{\rm max}$ between the coma and tail can be explained to some extent by modeling solutions assuming an identical composition and (for FAs) identical monomer size distributions with the same mean radius $r_{\rm mean}$ but different particle size distributions in the coma and tail. Both solutions have their shortcomings in terms of quality of fit with the observations and/or physical plausibility, which can essentially be overcome by assuming a mixture of FA and AD particles in the coma and the tail.
\paragraph{Solution~1 -- FA particles only.}
The observed differences in $P_{\rm max}$ and $\alpha_{\rm max}$ can in principle be reproduced by setting the parameters of the particle size distribution to $n_{\rm coma}=2$ and $c_{\rm coma}=0.6~\mu$m in the coma and to $n_{\rm tail}=5$ and $c_{\rm tail}=0.1~\mu$m in the tail, where $c_{\rm coma}$ and $c_{\rm tail}$ denote the smallest particle size in the distribution, respectively. The monomer size distribution is identical in the coma and tail with $r_{\rm mean}=137.5~\mu$m. The corresponding simulation results obtained for the coma and tail are shown in Fig.~\ref{fig:DoLP_modeled_sol1} for an identical composition of 50:50 olivine-carbon, respectively (only negligible changes of the DoLP occur when the composition is varied across the range between 30:70 and 70:30). In this parameter setting, the differences in $P_{\rm max}$ and $\alpha_{\rm max}$ between the coma and tail are caused only by the different values of $n$ and the omission of small agglomerates with $R_g<0.6~\mu$m in the coma. For both the coma and tail, this setting yields realistic $P_{\rm max}$ values, whereas $\alpha_{\rm max}$ is too small by about $10^\circ$--$20^\circ$. In conclusion, we were not able to identify an MSTM5 solution with identical composition and $r_{\rm mean}$ that reproduces the DoLP phase curves of the coma and tail simultaneously -- a problem that could not occur for previous modeling studies that took into account cometary comae only.

A possible way out of the limitation of underestimated $\alpha_{\rm max}$ values is to add AD particles to our model setting \citep[e.g.,][]{Halder2021,Halder2023}.
\paragraph{Solution~2 -- AD particles only.}
We examined to which extent it would be possible to alleviate the described modeling deficiency by adding those particles to our simulation (Sect.~\ref{sec:PolarizationModeling}), as they typically imply larger $\alpha_{\rm max}$ values than FAs for the same composition and size distribution. Using the software package REST \citep{Halder2022} that partially builds upon the DDSCAT framework \citep{Draine1994,Draine2013}, we simulated the DoLP phase curve of a multidisperse set of AD particles as described in Sect.~\ref{sec:PolarizationModeling}. We assumed the same compositions and particle size distributions as for solution~1, with 50:50 olivine-carbon and $n=2$ and $c=0.6~\mu$m in the coma and $n=5$ and $c=0.1~\mu$m in the tail (Fig.~\ref{fig:DoLP_modeled_sol2}). For the coma, the simulated value of $P_{\rm max}$ is too low while $\alpha_{\rm max}$ is too large. In contrast, the simulated tail DoLP coincides with the observed $P_{\rm max}$ and $\alpha_{\rm max}$ values.

It is interesting to note that according to Fig.~\ref{fig:ModelParameters3} the positive polarization branch of the coma and tail can be reproduced with an identical 30:70 olivine-carbon mixture, $c=0.1~\mu$m, and $n_{\rm coma}=2$ and $n_{\rm tail}=5$. However, the lack of a negative polarization branch of the simulated DoLP phase curve of the coma is not consistent with our (limited) observations at small phase angles.
\paragraph{Solution~3 -- mixture of particle morphologies.}
An important finding of the in-situ study by \citet{Guettler2019} using Rosetta data was that both FA and more compact ``solid'' particles were found in the coma of 67P/Churyumov-Gerasimenko. From the modeling perspective, it is thus a straightforward step to assume a mixed solution, i.e., a weighted average of DoLP phase curves computed with MSTM5 and DDSCAT, respectively, provided that the compositions and size distributions are the same for both particle types (Fig.~\ref{fig:DoLP_modeled_sol3}). A 50:50 FA vs.\ AD mixture yields a modeling result that coincides very well with our coma measurements. The same mixture is also reasonably consistent with our tail measurements, but a 25:75 mixture appears to yield a still better fit, where the relatively high measurement errors do not provide a reliable constraint.
\section{Discussion}
\label{sec:Discussion}
\subsection{Observations}
\label{sec:DiscussionObservations}
The presented polarization observations of the coma and tail of comet C/2023~A3 (Tsuchinshan-ATLAS) illustrate that wide-field polarimetric image data can be obtained in the positive polarization branch for moderately bright comets with a small instrument of short aperture ratio in combination with an industrial-grade polarization camera. In this way, we obtained imaging polarimetric observations of the tail of C/2023~A3 across a broad range of phase angles. In contrast, the negative polarization branch is less easily accessible due to the significantly lower DoLP, requiring a higher SNR of the polarimetric imaging sensor that might be achievable by an active cooling unit.

With a maximum DoLP of $P_{\rm max}=0.34$ obtained at $\alpha_{\rm max}\approx 100^\circ$, and $\alpha_{\rm inv}=20.5^\circ$, our coma measurements are consistent with those of \citet{Lim2025}, who obtained $P_{\rm max} \approx 0.31$ in the Rc band with $\alpha_{\rm max}$ corresponding to $90^\circ$--$95^\circ$ and $\alpha_{\rm inv} \approx 20^\circ$. Their estimate of $\alpha_{\rm max}$ is probably more reliable than ours as their phase angle coverage extends up to $123^\circ$. Our measurements of the tail reveal a still nearly two times higher $P_{\rm max}$ value of about $0.64$ obtained at $\alpha_{\rm max}$ slightly above $80^\circ$. A qualitatively similar increase of the DoLP from the coma towards the tail is apparent in polarimetric image data of the comet C/2010~E6 (STEREO) \citep{Nezic2022}. The estimated phase angle of inversion of $\alpha_{\rm inv}=20.5^\circ$ lies well within the range of values found for a variety of other comets \citep[e.g.,][]{Kiselev2015}.

Regarding the negative polarization branch of C/2023~A3, there is only limited coverage by our observational data. The two broadband measurements at phase angles of $16^\circ$--$17^\circ$ indicate absolute DoLP values of $0\geq P\gtrapprox -0.03$ in the central coma (Table~\ref{tab:Observations1}). According to Fig.~\ref{fig:ErrorAnalysisResults}, the corresponding DoLP values listed in Table~\ref{tab:Observations1} are lower limits to the actual negative DoLP because due to the low coma brightness, the SNR$_F$ level in the coma was found to be around 70 after binning. These measurements are consistent with the EMPOL observation of Mar~06, 2025, indicating a negative DoLP of $P\leq -0.0073$ in the SLOAN r~band at $14.1^\circ$ phase angle (Table~\ref{tab:Observations2}).

An image of comet C/2023~A3 that we took with a cooled color (RGB) CMOS camera and the 30/135~mm refractor on Oct~22, i.e., just one day before our first polarimetric imaging observation, does not show a green C$_2$ emission. Notably, we have been able to detect such a component easily with the same camera for several other, even much fainter comets, such as 13P/Olbers and 12P/Pons-Brooks, in 2024 (Fig.~\ref{fig:ComparisonToCometsWithC2Emission}). The post-perihelion spectrum of the coma of C/2023~A3 shown in Fig.~\ref{fig:SpectrumPostPerihelion_Sensitivity} clearly reveals a C$_2$ emission line. That is in contrast to the pre-perihelion spectrum acquired by \citet{Cambianica2025} on May~01, 2024, in which the C$_2$ emission line is barely visible, so that they classified C/2023~A3 as ``carbon-depleted''. Also the post-perihelion integrated C$_2$ flux, though, is lower than the continuum flux integrated over the green and red spectral domain where our cameras are most sensitive. This explains that the coma of C/2023~A3 does not appear green in the RGB image in Fig.~\ref{fig:ComparisonToCometsWithC2Emission}a. Furthermore, no gas tail is apparent in Fig.~\ref{fig:ComparisonToCometsWithC2Emission}a. These findings indicate a negligible contribution of line emission to the observed polarization.
\subsection{Modeling}
\label{sec:DiscussionModeling}
For all our obtained modeling solutions that are able to explain the observations reasonably well, an identical composition of the coma and tail and (in the case of FA particles) monomer size is assumed. In contrast, the parameters of the particle size distributions of the coma and tail are different in the way that the tail is dominated by smaller particles than the coma. A straightforward physical explanation for this difference in size distribution is as follows: Large (micrometer-sized) agglomerates freshly released from the nucleus into the coma are composed of smaller agglomerates ``glued'' to each other by frozen volatiles. At distances of more than about 1000~km from the nucleus, their transport from the coma into the tail is mainly driven by the solar radiation pressure \citep{Agarwal2024}. During their motion, the agglomerates are warmed up by solar radiation so that the volatiles sublimate and the large agglomerates decay into smaller constituents or even individual monomers that cause the observed polarization properties. \citet{Agarwal2024} showed that the acceleration of a particle induced by radiation pressure relative to its gravitational accelation increases with decreasing particle radius. This behavior implies that small particles are transported from the coma to the tail more efficiently than large particles, a mechanism that readily explains the predominance of small particles in the tail indicated by our modeling results.

According to model solution~1 (Fig.~\ref{fig:DoLP_modeled_sol1}), the coma is characterized by particles with $R_g>0.6~\mu$m and a power law exponent of $n=2$, whereas in the tail there is a stronger influence of small particles with sizes close to the Rayleigh limit, expressed by a large power law exponent of $n=5$ and $R_g>0.1~\mu$m. The assumed composition is an identical 50:50 olivine-carbon mixture in the coma and tail. A reasonable fit of the simulation result to our coma observations is obtained  for both the positive and the negative polarization branch, with $\alpha_{\rm max}$ being about $10^\circ$ too small, though. However, the simulated $\alpha_{\rm max}$ value of the tail is too small by about $20^\circ$ when compared to the observations.

For a mean monomer size of $r_{\rm mean}=137.5$~nm, the maximum gyration radius $R_g$ of our modeled FA and AD particles corresponds to $10.1~\mu$m and $2.0~\mu$m, respectively, with radii of the corresponding equal-volume spheres of $1.03~\mu$m and $1.71~\mu$m, respectively. At this point, one might think of adding large particles to the simulation, as laboratory measurements by \citet{Munoz2020} show that millimeter-sized cosmic dust analog particles exhibit high values of $\alpha_{\rm max}$ between $110^\circ$ and $140^\circ$. However, the total geometric cross-section of an ensemble of particles obeying a power-law size distribution with $n>3$ has been shown to be largely independent of the occurrence of particles of several micrometers size based on analytical calculations as far as scattering of visible light is concerned \citep{Zubko2020}. This means that our tail simulations with $n=5$ would remain unaffected by an addition of large particles. But even for $n \leq 3$, their simulations indicate that including particles of comet-like composition with size parameters $x=2 \pi r/\lambda>24$ (corresponding to radii of $r>2.3~\mu$m at $\lambda=600$~nm wavelength) will at the most lead to a slight underestimation of $P_{\rm max}$ by typically less than $0.01$, where the strongest effect occurs at phase angles beyond $80^\circ$ that are hardly covered by our observations. Consequently, the contribution of such large particles to the observed DoLP will only be non-negligible in the presence of mechanisms that remove small (sub-micrometer) particles from the ensemble \citep{Zubko2020}. Furthermore, it has been shown by \citet{Kolokolova2015} in their numerical study of the polarization behavior of large rough spheroidal particles that a negative polarization branch, as detected for the coma of C/2023~A3, only exists for size parameters below $30$. In addition to that, both MSTM5 and DDSCAT are not appropriate for modeling the light scattering behavior of particles that are much larger than the wavelength of light, as for this size range it would be required to incorporate principles of geometric optics \citep{Markkanen2016,Markkanen2018}. It would certainly be interesting to investigate quantitatively the effects of an absence of sub-micrometer particles on the ensemble's polarization behavior and add specific modeling techniques for large particles, but these questions are beyond the scope of this paper.

According to model solution~2 (Fig.~\ref{fig:DoLP_modeled_sol2}), the DDSCAT-based simulation result yields a too low $P_{\rm max}$ and a too large $\alpha_{\rm max}$ value for the coma with the same composition and particle size distribution as for MSTM5-based solution~1. In contrast, the simulated DoLP phase curve of the tail reproduces our observations accurately. When assuming size distributions for both the coma and tail that comprise particle sizes down to $0.1~\mu$m, the DDSCAT-based simulations realistically reproduce the respective positive polarization branches when assuming a 30:70 olivine-carbon mixture (Fig.~\ref{fig:ModelParameters3}). The main deficiency of such a purely DDSCAT-based solution, though, is the absence of a negative polarization branch of the coma. Furthermore, the atomic force microscope analyses of dust particles of comet 67P/Churyumov-Gerasimenko on-board the Rosetta spacecraft described by \citet{Guettler2019} indicate that it would be unrealistic to assume an absence of FAs in the coma, as this type of structure was clearly found to be abundant on micrometer and sub-micrometer scales. Such direct observations are not available for the tail, though.

These considerations took us to a model setting involving mixtures of FA and AD particles (solution~3, Fig.~\ref{fig:DoLP_modeled_sol3}). A 50:50 mixture yields a good fit to our coma observations, but the value of $\alpha_{\rm max}$ modeled for the tail is still too low by about $10^\circ$ although the overall correspondence with our tail observations is reasonable. Increasing the AD fraction to 75\% improves the model fit for the tail.

If this difference in the AD fraction was real (which cannot be confirmed definitely due to the limited accuracy of our tail observations), it could be due to a dynamical effect to be explored in more detail that drives porous particles out of the tail more efficiently than compact particles of similar size. Weathering of sub-micrometer and micrometer-sized FA particles towards AD morphology by, e.g., solar radiation or high-energy solar wind particles, during their motion from the coma into the tail is probably not a viable mechanism, as can be shown by an elementary kinematic consideration. At 1~AU solar distance and beyond the gravitational influence of the nucleus, the largest modeled FA particles with $R_g \approx 10~\mu$m and equal-volume sphere radius of $1~\mu$m would experience a solar radiation force of roughly $10^{-15}$~N \citep{Agarwal2024}. Assuming a density of $3 \times 10^3$~kg~m$^{-3}$ typical of silicate minerals implies a time of about $15$~hours until the particle reaches a distance of $10^5$~km from the nucleus. Significant changes in the particle morphology are unlikely to occur on such short timescales. However, compact weathered particles of AD morphology were possibly formed on the surface of the nucleus \citep{Kwon2019} prior to the formation of the tail and had already been released into the tail by the time of our observations, while subsurface FA particles of higher porosity were released later and still dominated the coma.

Alternative explanations of the tail's large $P_{\rm max}$ value include minor variations in the monomer size and/or the imaginary part of the refractive index of the FA particles \citep{Kimura2006}. Similarly, the modeled $P_{\rm max}$ value could also be increased by choosing non-identical size ranges for FA and AD particles in the tail. More observational data of the dust particles in cometary tails are needed, though, for a clear identification of the relevant physical parameters.

It is not strictly necessary for our model, though, to invoke such processes in order to provide a reasonable simultaneous model fit to our coma and tail DoLP data, as also with identical FA vs.\ AD fractions in the coma and tail the modeled tail DoLP largely remains within the measurement error intervals.
\section{Conclusions and future work}
\label{sec:Conclusions}
In this study we analyzed the size and structure of the dust particles in the coma and tail of comet C/2023~A3 (Tsuchinshan-ATLAS) based on wide-field polarimetric imaging. We measured the DoLP of the coma and tail across a broad range of phase angles between $16.4^\circ$ and $85.5^\circ$. Based on a Monte-Carlo simulation, we quantified both random and systematic errors of our DoLP measurements in dependence on the image SNR. With a maximum DoLP of $0.34$ in the coma, comet C/2023~A3 belongs to the class of highest-polarization comets, and the DoLP of the tail even becomes as high as $0.64$. The phase angle of the DoLP maximum corresponds to a (fairly uncertain) value of $104^\circ$ in the coma and to $86^\circ$ in the tail. We detected the presence of a negative polarization branch with a minimum negative DoLP larger than approximately $-0.03$.

Our modeling results suggest an identical 50:50 olivine-carbon composition and the same monomer size distribution with $r_{\rm mean}=137.5~\mu$m in the coma and tail when FAs are considered as the only morphological type of dust particles present. However, the simulated DoLP phase curves do not provide a good fit to our observations under that assumption, whereas admixing AD in fractions of at least 50\% into the coma and tail particles yields a reasonably accurate simultaneous model fit to our coma and tail data. The assumption of both FAs and AD occurring in the coma is plausible according to data from the Rosetta spacecraft \citep{Guettler2019}, whereas no similar in-situ observations are available for any cometary tail. A possibly higher fraction of AD particles in the tail may be due to weathering processes acting on the FA particles and transforming a part of them into AD morphology.

For the coma, we found a moderately steep particle size distribution with $n=2$ and an absence of particles with radii below $0.6~\mu$m. For the tail, our simulations indicate a steep size distribution with $n=5$ and $0.1~\mu$m cutoff, indicating a predominance of individual monomers and small AD particles with sizes close to the Rayleigh limit. This finding is consistent with smaller particles being transported from the coma into the tail more efficiently by the solar radiation pressure than larger particles.

Future work will involve wide-field imaging polarimetric observations of the tails of different types of comets, preferentially with more sensitive sensors allowing for the use of photometric filters covering a broad spectral range in the visible and near-infrared domain while maintaining a high SNR. For such observations, broad phase angle coverage will be of central importance, not only at large phase angles in order to accurately determine the phase angle of maximum DoLP but also at small phase angles for precisely estimating the minimum DoLP and the phase angle at which it occurs.  For a better interpretability of the observations, a further refined modeling analysis might also consider the roles of other physical characteristics, such as particle porosity and, following the statement of \citet{Guettler2019} that ``monomers in nature are not perfectly spherical and have rough surfaces'', rough-surfaced monomers of non-spherical shape. Simultaneous modeling of the coma vs.\ tail dust particles will reveal whether the corresponding differences found for C/2023~A3 are of general nature or depend on the comets' individual properties.
\section*{Data availability}
The polarimetric image data analyzed in this study and the modeling results are available at \url{https://doi.org/10.5281/zenodo.18156295} for download.
\section*{Acknowledgements}
The authors would like to thank Vikrant Agnihotri for providing the post-perihelion spectrum shown in Fig.~\ref{fig:SpectrumPostPerihelion_Sensitivity}. C.~W. acknowledges partial funding by the German Academic Exchange Service (DAAD) Bilateral Exchange of Academics program (project 0002857187) and friendly support by the ARIES Observatory staff, in particular Dr.\ Santosh Joshi. Work at PRL is funded by the Dept.\ of Space, Govt.\ of India. The authors thank the staff at the Mount Abu Observatory for their support in the observations. The authors would furthermore like to thank an anonymous reviewer for the insightful comments and suggestions.
\bibliography{pol}
\onecolumn
\begin{appendix}
\section{Post-perihelion spectrum of comet C/2023~A3}
\begin{figure*}[hb]
\centering
\includegraphics[width=0.8\textwidth]{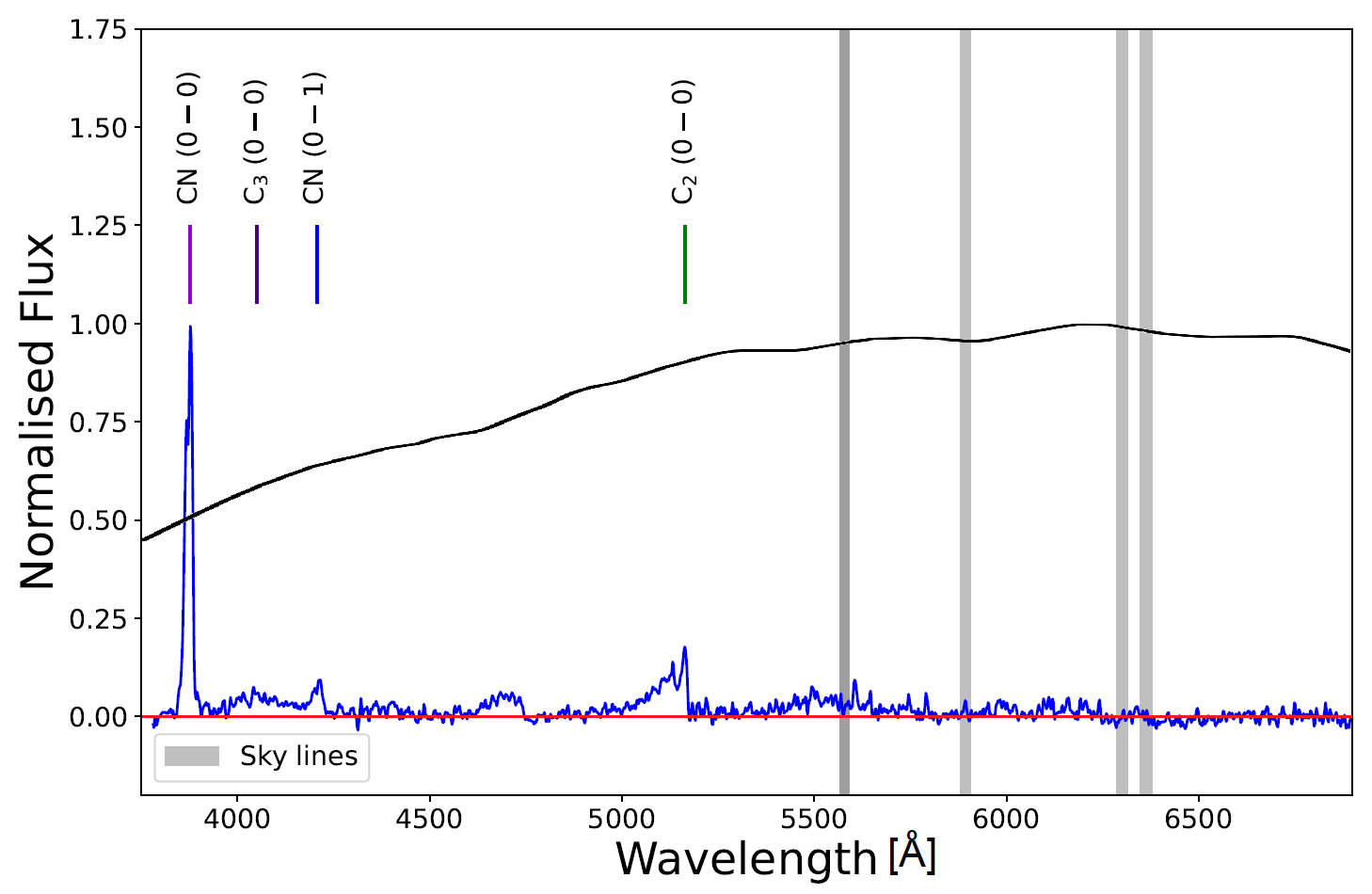}
\caption{Post-perihelion spectrum of the coma of comet C/2023~A3 acquired by Vikrant Agnihotri on Nov~02, 2024, at 13:13 UT, with a phase angle of $57.4^\circ$, an airmass of 1.54, and an exposure time of 900~s. The black curve denotes the relative sensitivity of the DZK~33UX250 polarization camera \citep{TIS2019}.}
\label{fig:SpectrumPostPerihelion_Sensitivity}
\end{figure*}
\newpage
\section{Comparison of comet C/2023~A3 to comets with C$_2$ emission}
\begin{figure*}[hb]
\centering
\includegraphics[width=1.0\textwidth]{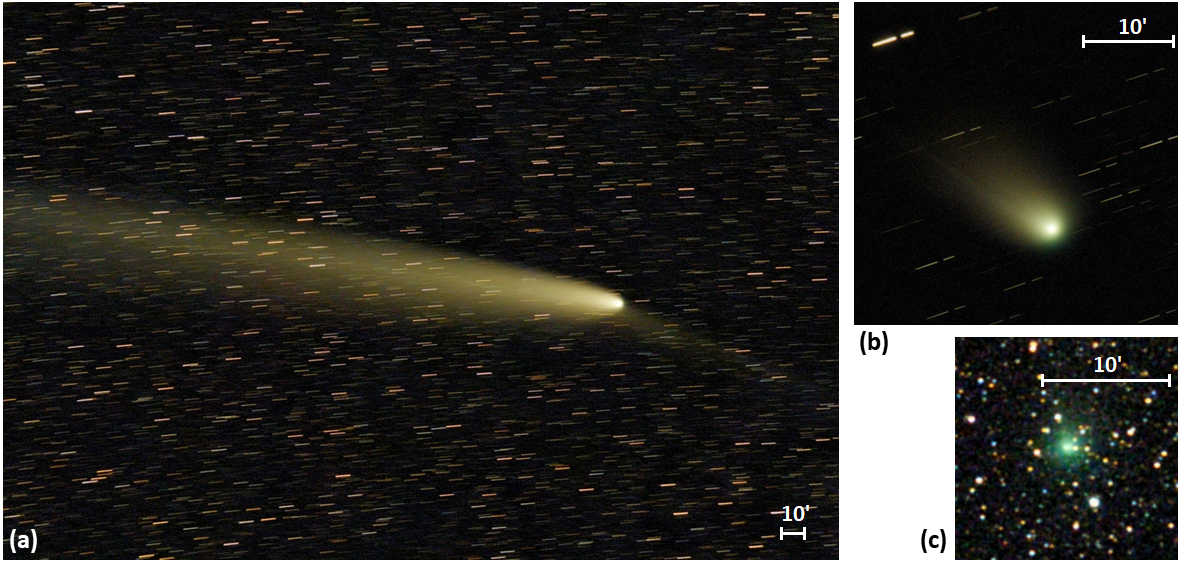}
\caption{Comet C/2023~A3 in comparison to other comets based on telescopic images acquired with the same cooled CMOS color camera. Nonlinear contrast stretching was applied. (a)~Comet C/2023~A3 (Tsuchinshan-ATLAS) on Oct~22, 2024, taken with 30/135~mm refractor. No green C$_2$ emission is detectable. (b)~Comet 13P/Olbers on Jul~29, 2024, taken with 150/600~mm Newton reflector. Part of the coma shows C$_2$ emission. (c)~Comet 12P/Pons-Brooks on Jan~27, 2024, taken with 30/135~mm refractor. The coma is dominated by C$_2$ emission.}
\label{fig:ComparisonToCometsWithC2Emission}
\end{figure*}
\newpage
\section{Shapes of the simulated agglomerates}
\begin{figure*}[hb]
\centering
\includegraphics[width=0.62\textwidth]{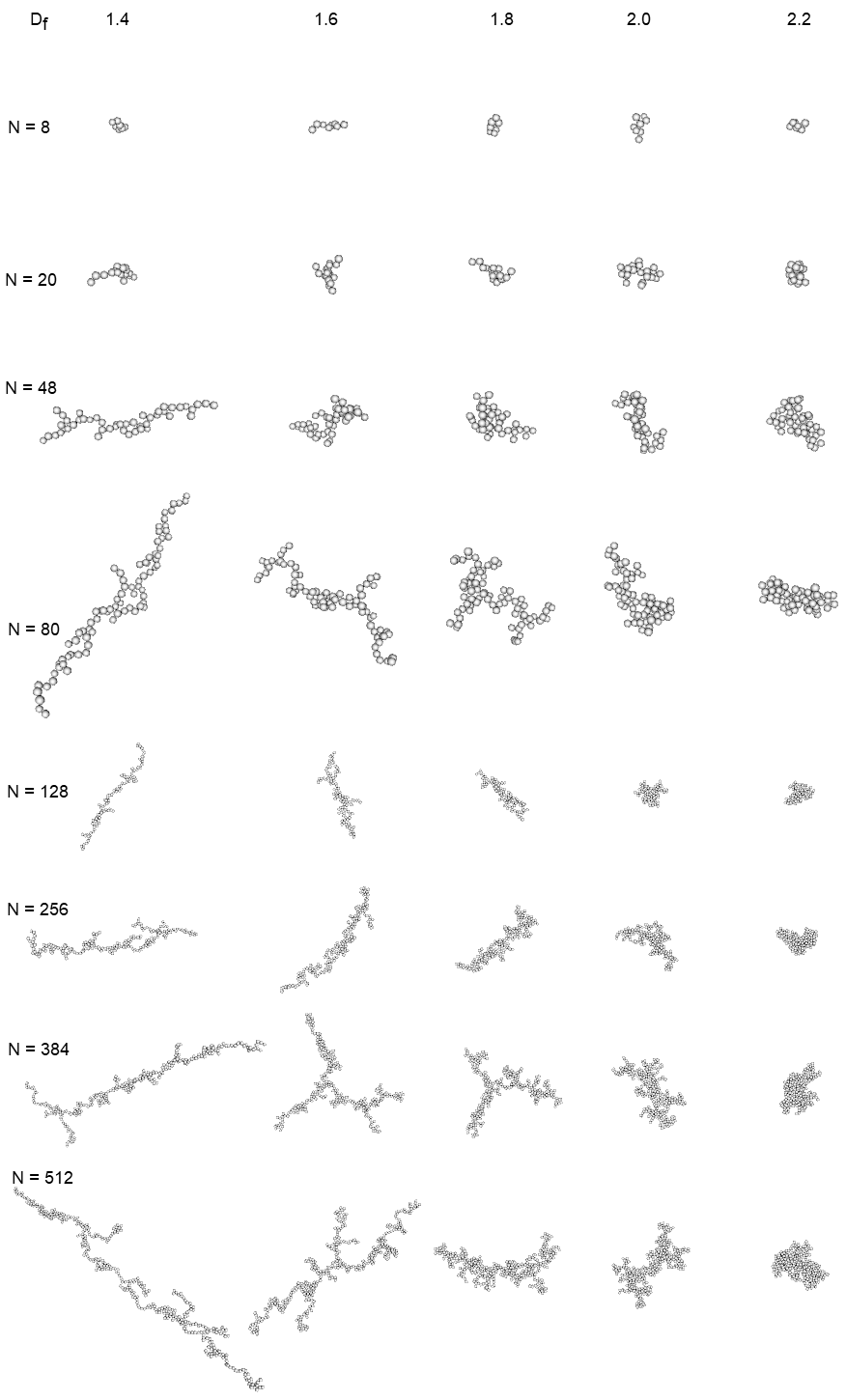}
\caption{Renderings of the fractal-shaped agglomerates used in our modeling study. The number $N$ of monomers per agglomerate corresponds to $8$, $20$, $48$, $80$, $128$, $256$, $384$ and $512$. The fractal dimension $D_f$ corresponds to $1.4$, $1.6$, $1.8$, $2.0$ and $2.2$, respectively.}
\label{fig:Agglomerates}
\end{figure*}
\newpage
\section{Influence of different parameters on the modeling results}
\begin{figure*}[hb]
\centering
\includegraphics[width=0.45\textwidth]{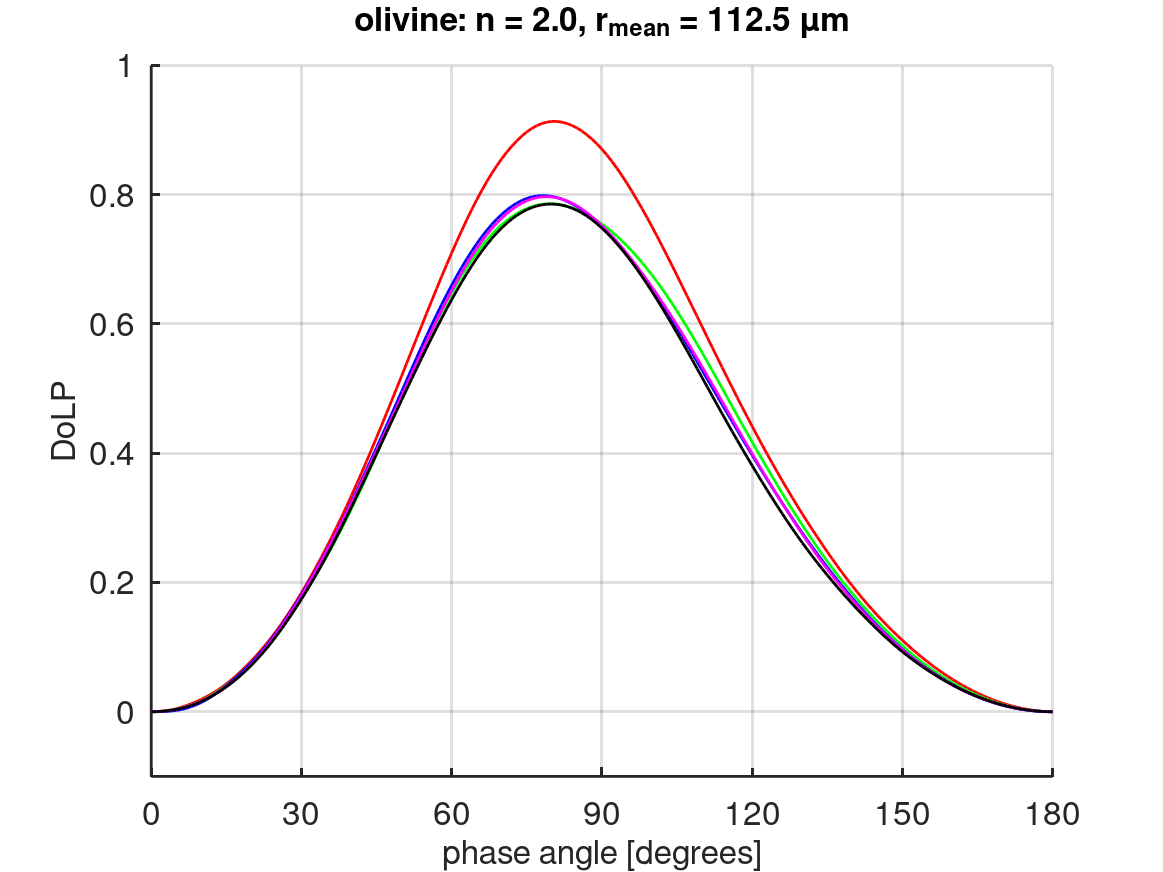}
\includegraphics[width=0.45\textwidth]{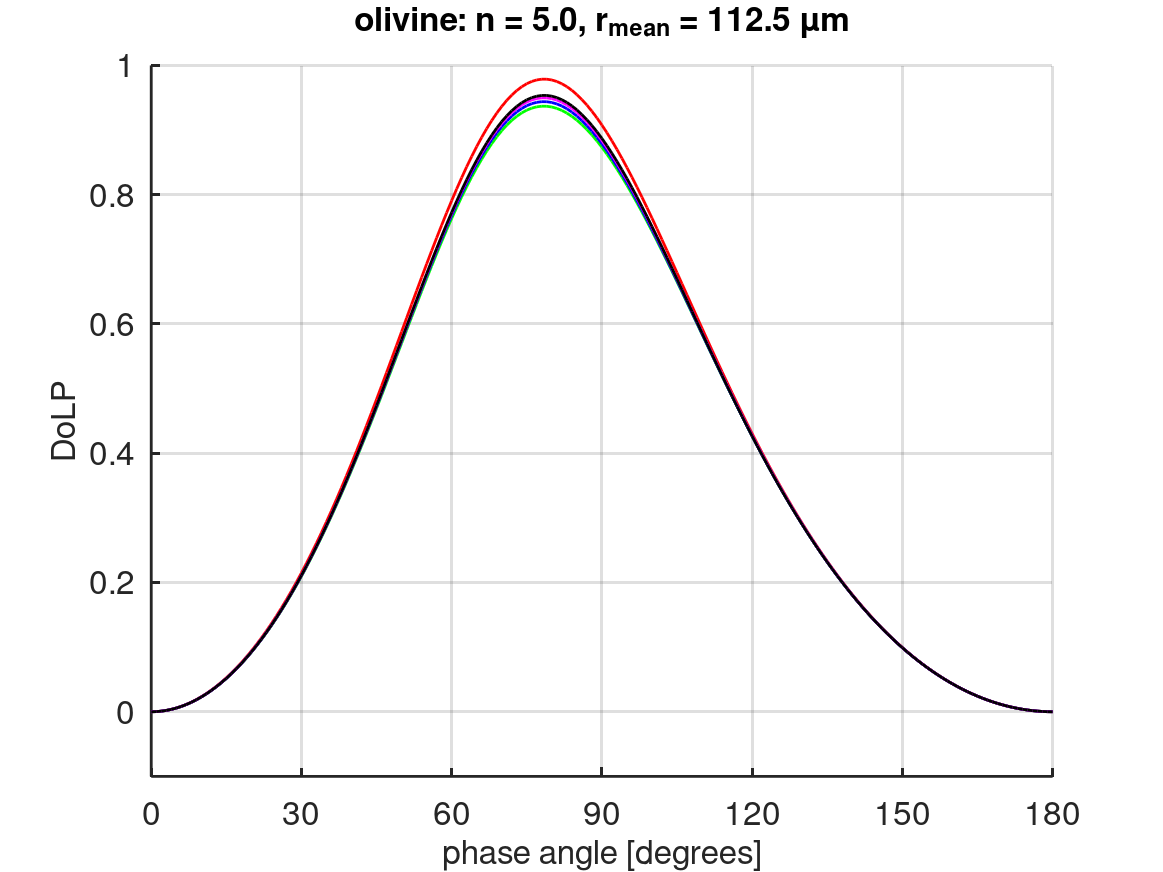}\\[1ex]
\includegraphics[width=0.45\textwidth]{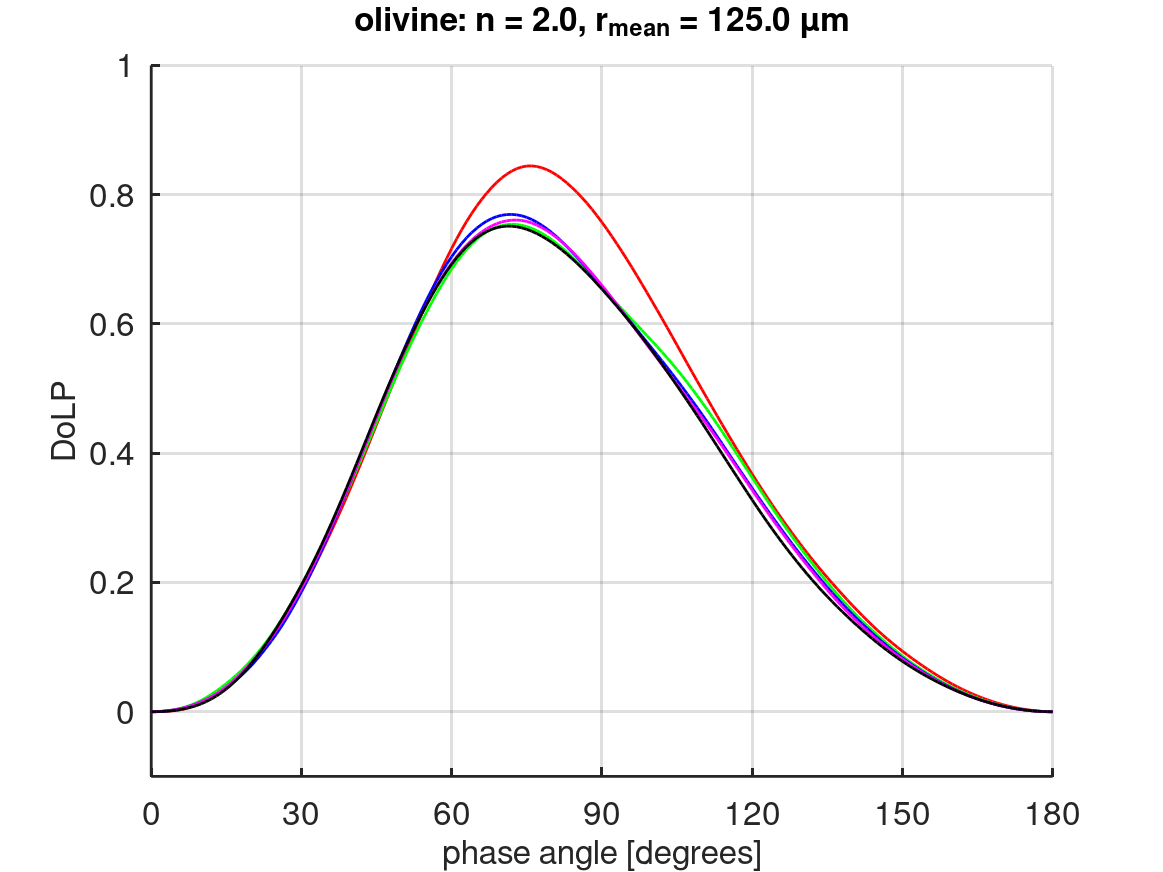}
\includegraphics[width=0.45\textwidth]{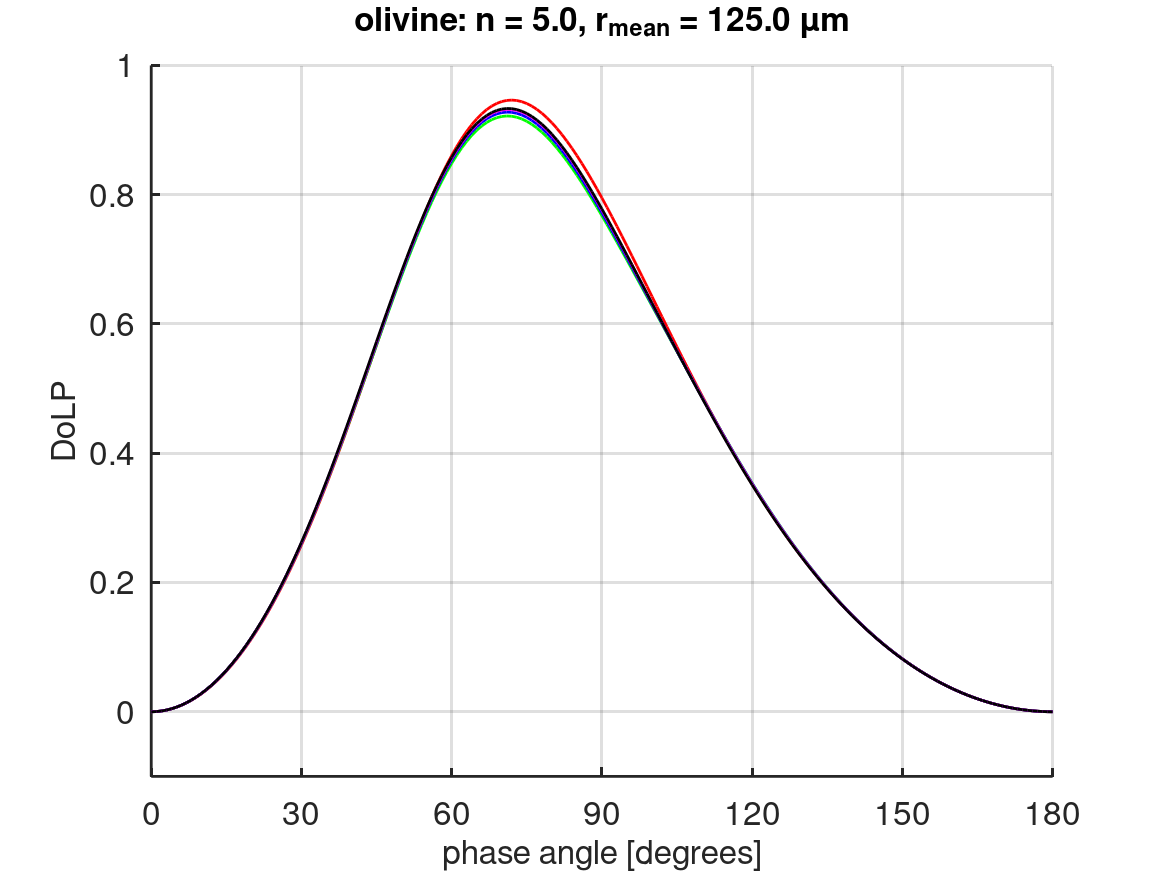}\\[1ex]
\includegraphics[width=0.45\textwidth]{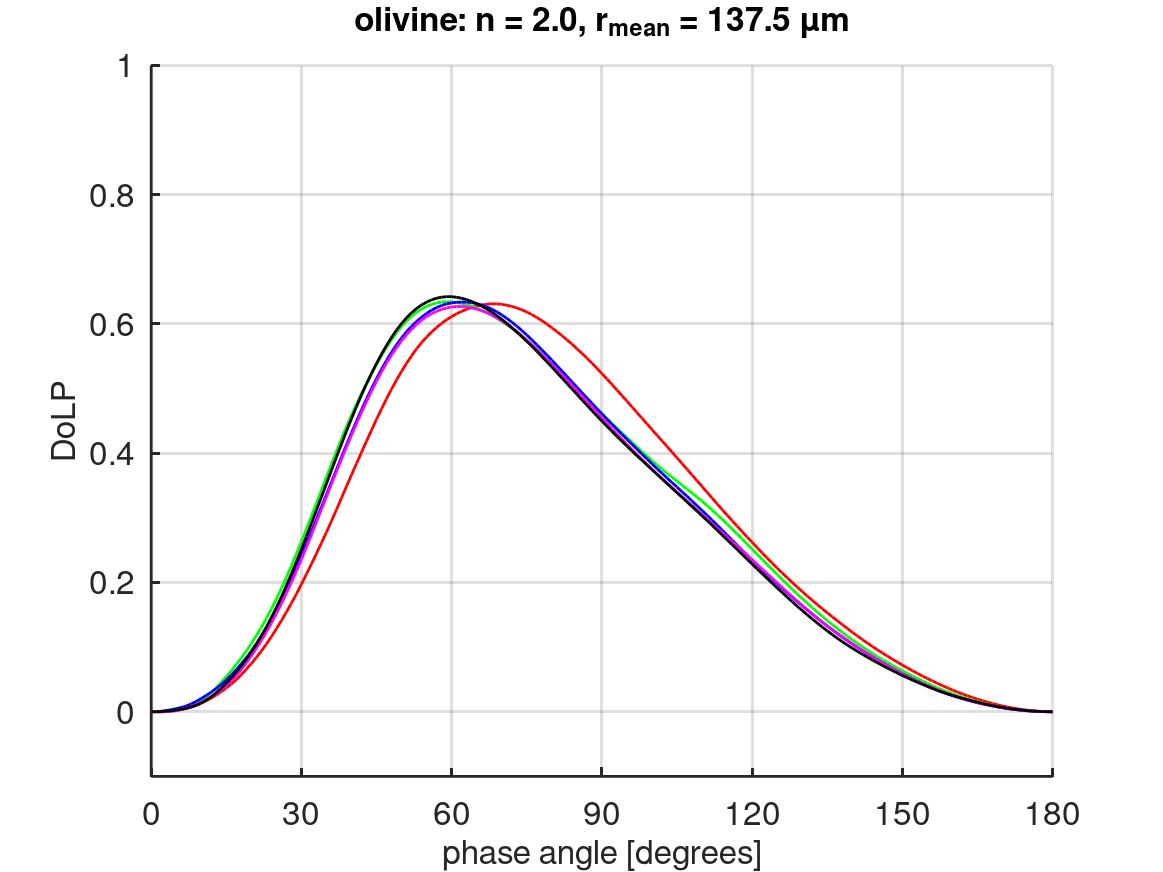}
\includegraphics[width=0.45\textwidth]{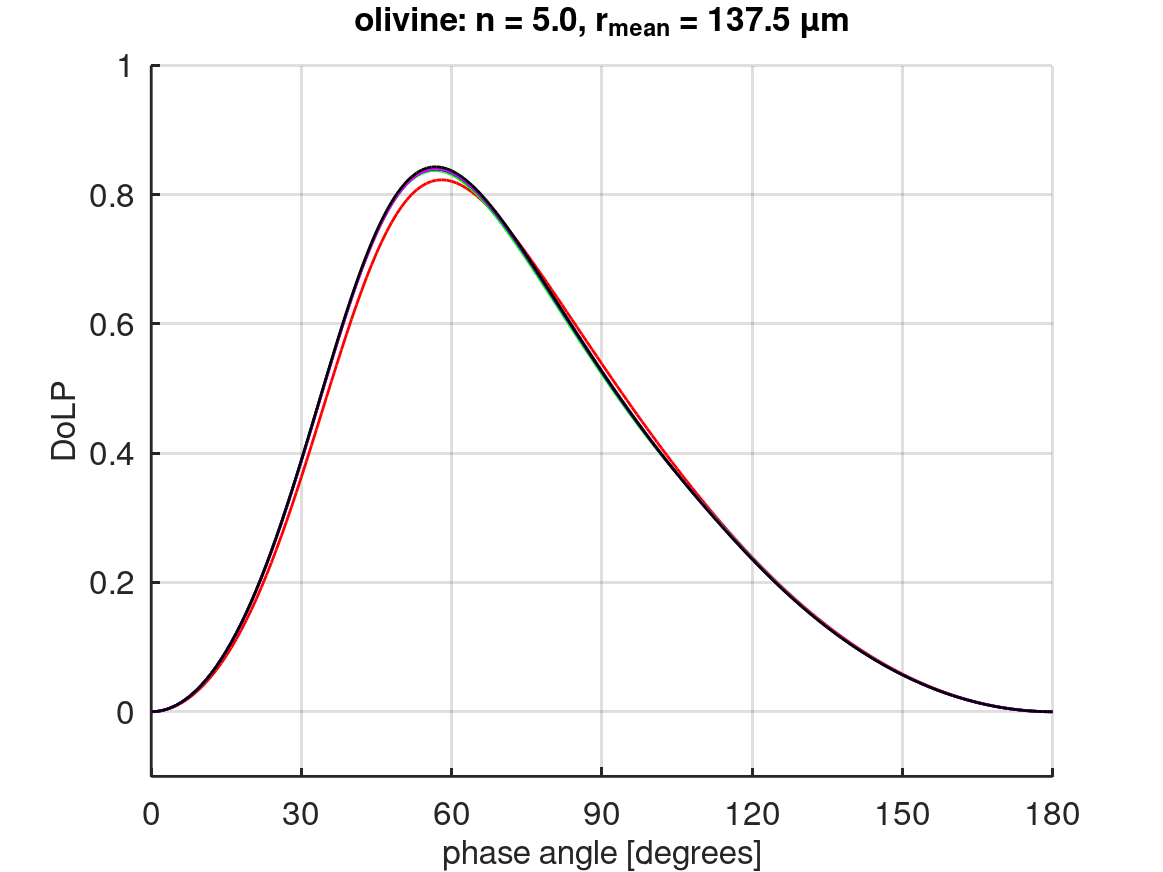}
\caption{MSTM5-based DoLP phase curves of pure olivine for mean monomer radii of $112.5$, $125$ and $137.5$~nm and power-law exponents of the particle size distribution of $2$ and $5$. Red, green, blue, magenta and black curves correspond to $D_f$ values of $1.4$, $1.6$, $1.8$, $2.0$ and $2.2$, respectively.}
\label{fig:ModelParameters1}
\end{figure*}
\begin{figure*}[hb]
\centering
\includegraphics[width=0.45\textwidth]{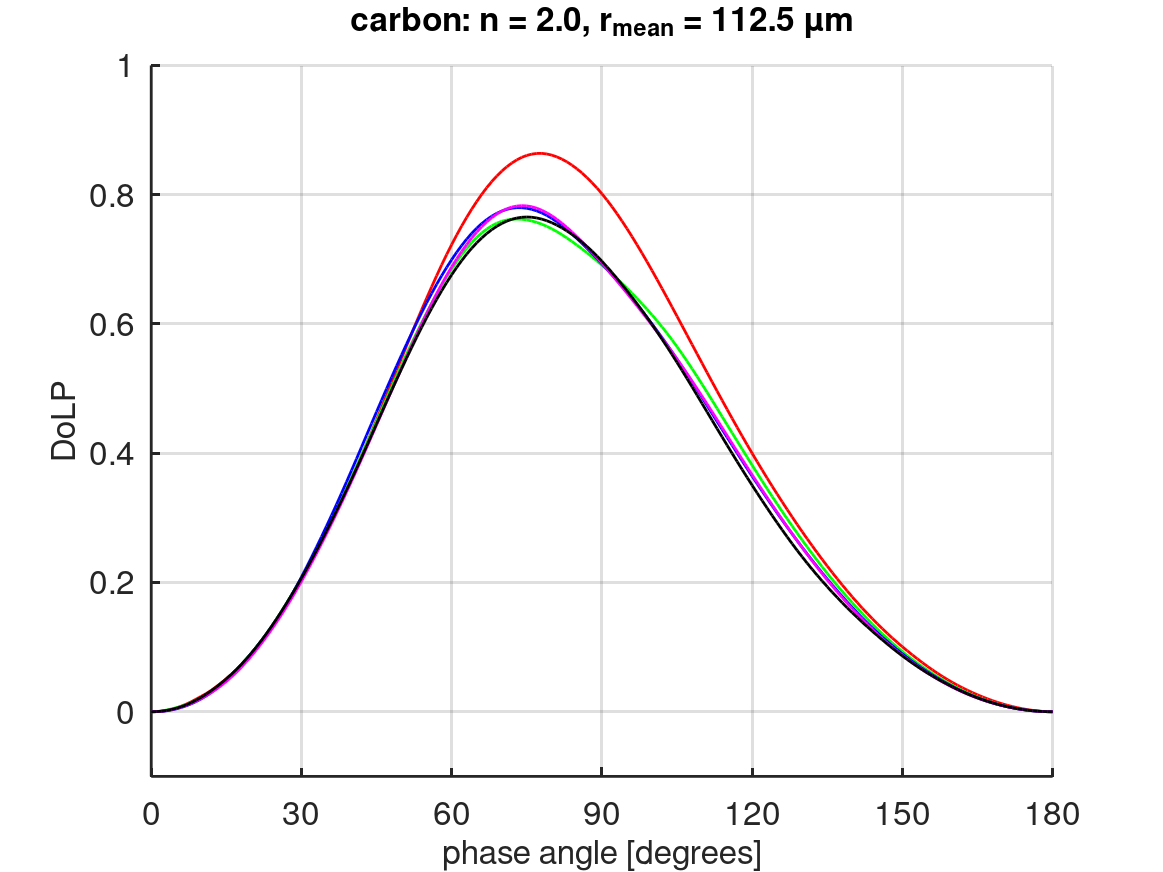}
\includegraphics[width=0.45\textwidth]{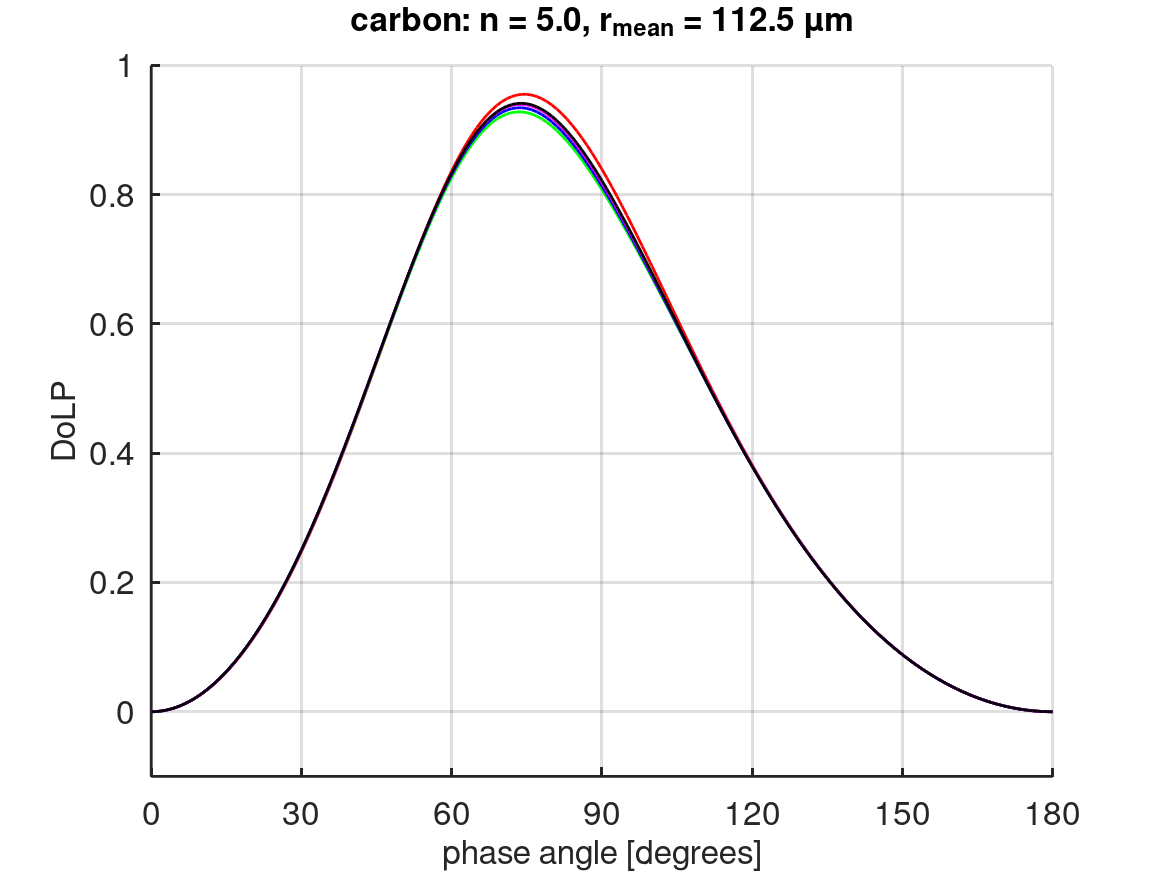}\\[1ex]
\includegraphics[width=0.45\textwidth]{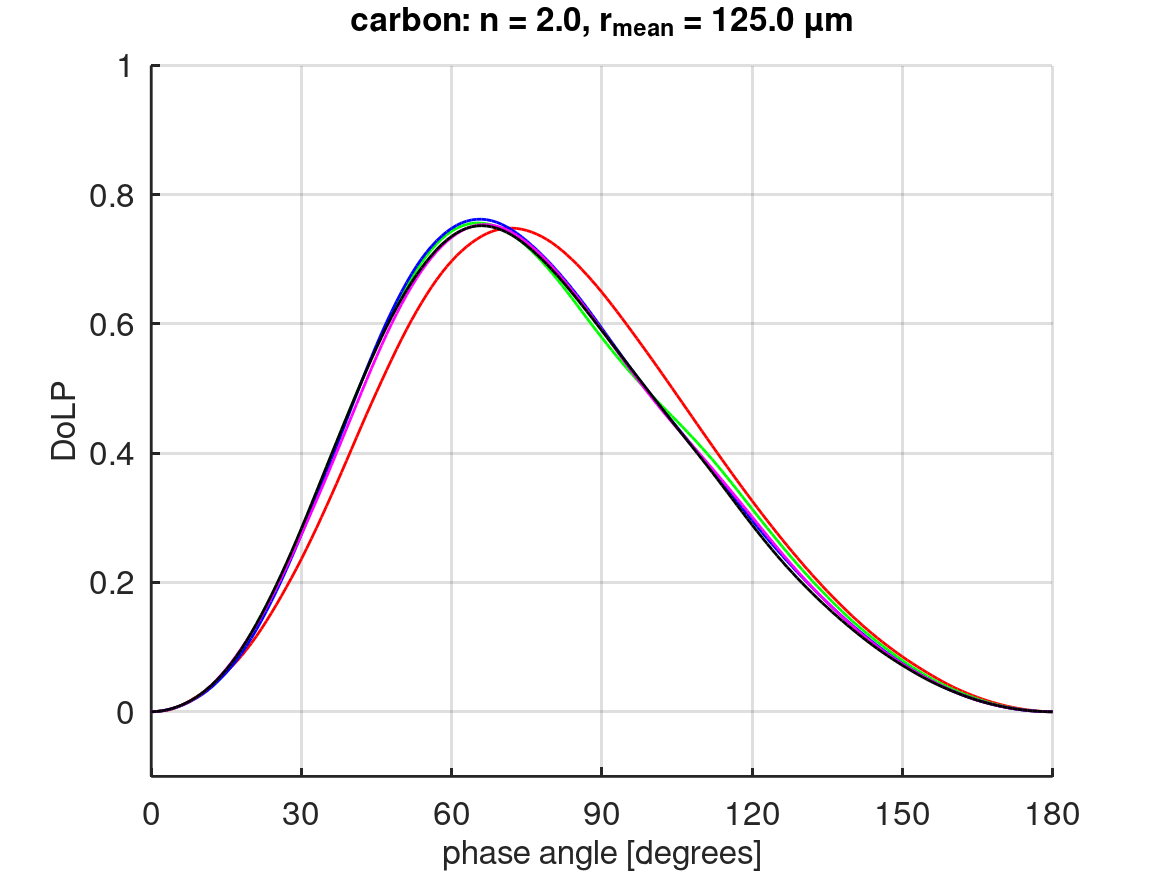}
\includegraphics[width=0.45\textwidth]{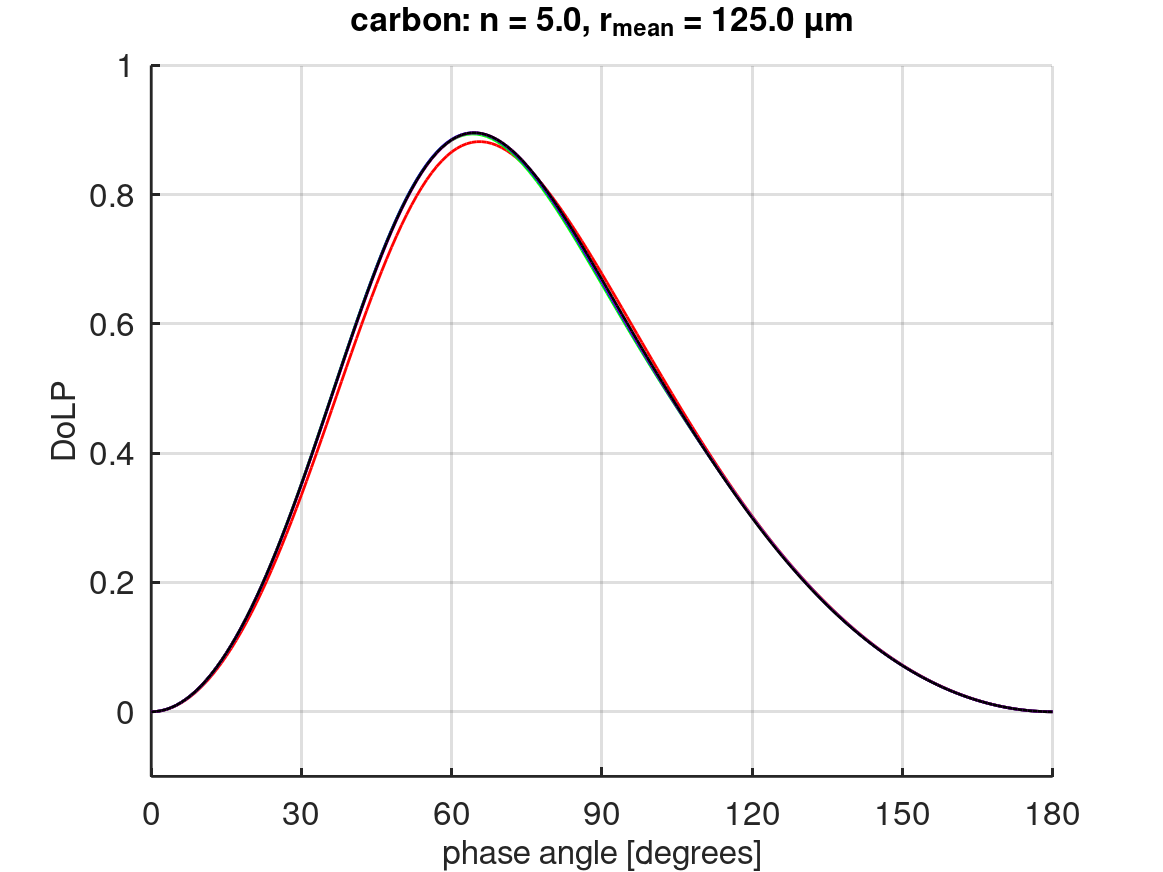}\\[1ex]
\includegraphics[width=0.45\textwidth]{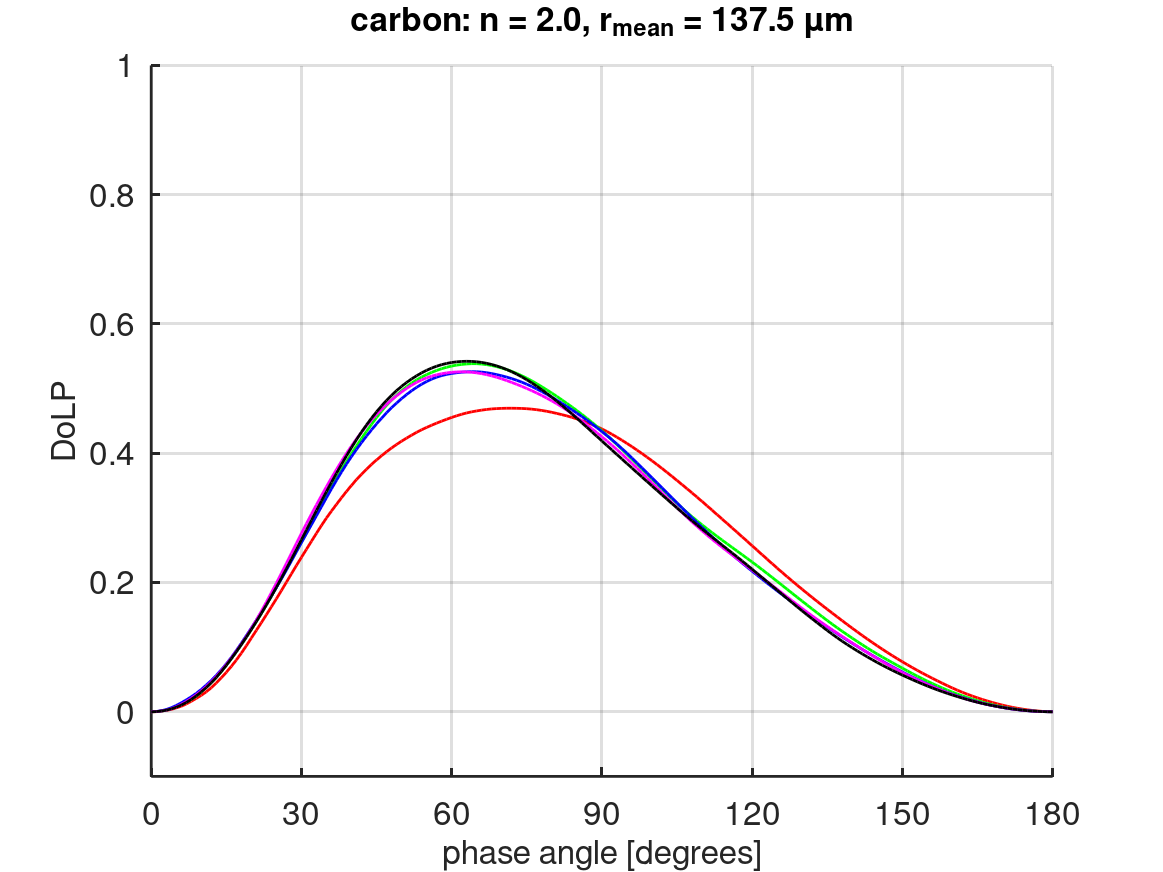}
\includegraphics[width=0.45\textwidth]{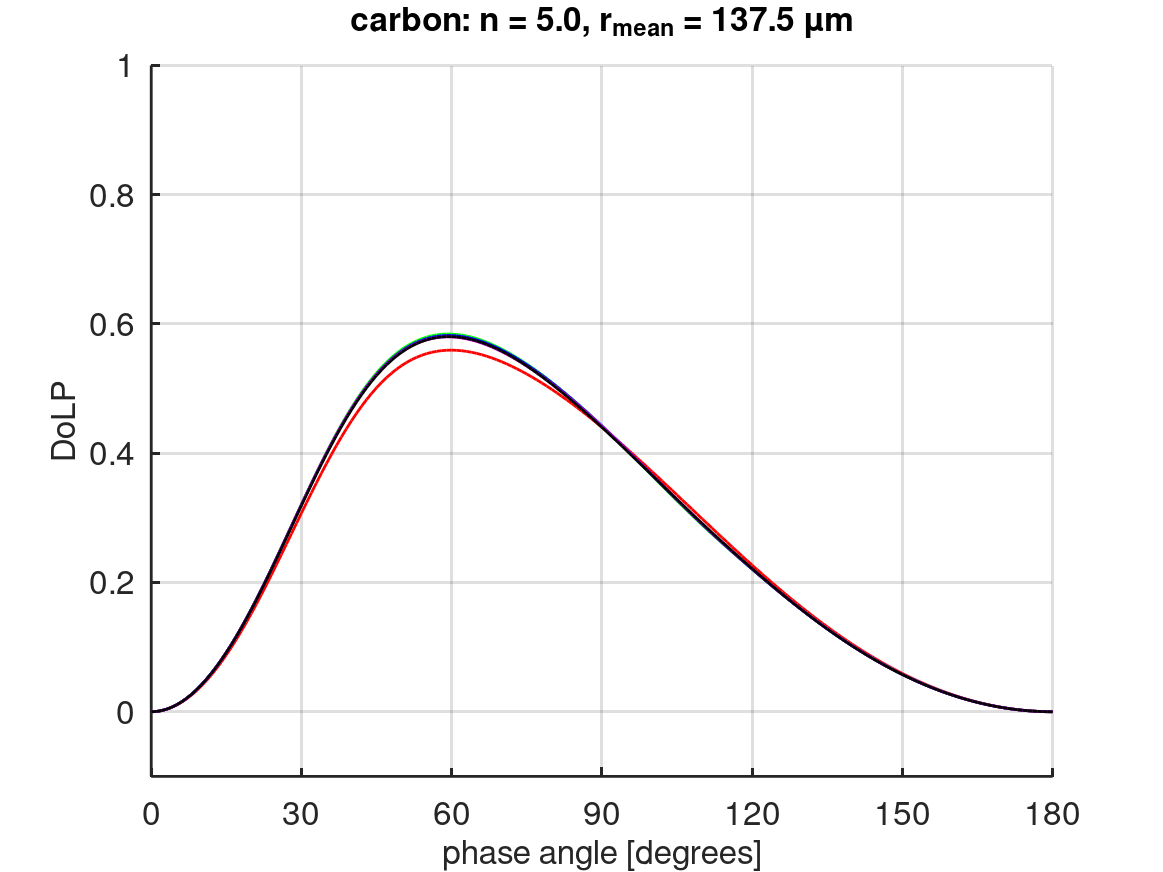}
\caption{MSTM5-based DoLP phase curves of pure amorphous carbon for mean monomer radii of $112.5$, $125$ and $137.5$~nm and power-law exponents of the particle size distribution of $2$ and $5$. Red, green, blue, magenta and black curves correspond to $D_f$ values of $1.4$, $1.6$, $1.8$, $2.0$ and $2.2$, respectively.}
\label{fig:ModelParameters2}
\end{figure*}
\begin{figure*}[hb]
\centering
\includegraphics[width=0.45\textwidth]{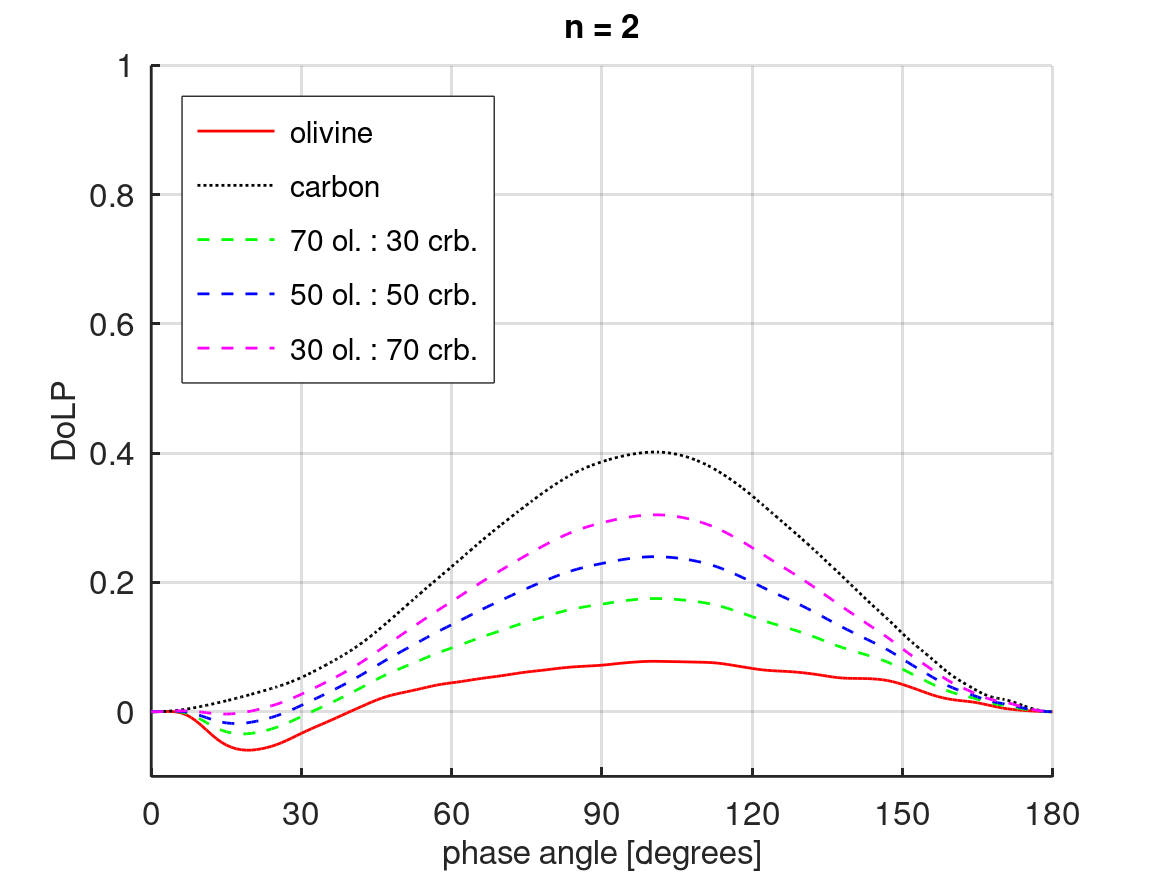}
\includegraphics[width=0.45\textwidth]{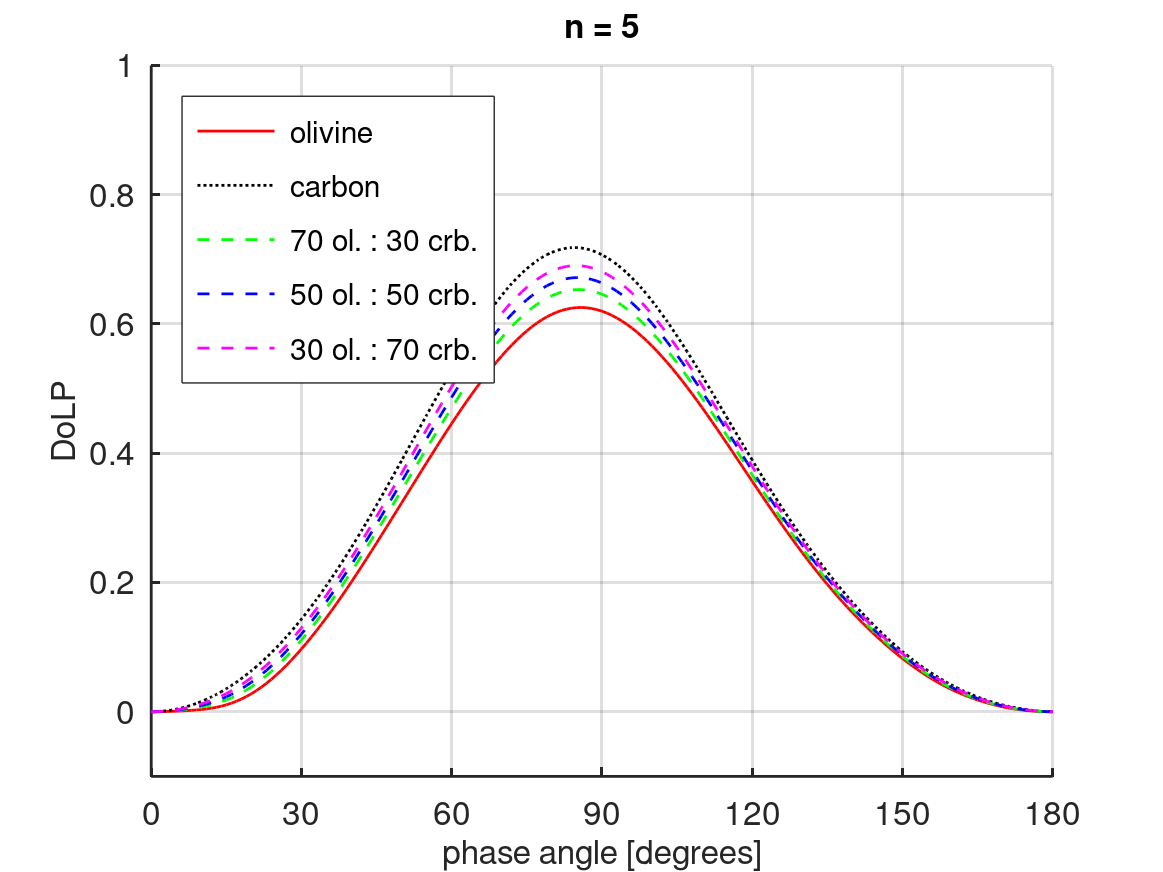}
\caption{DDSCAT-based DoLP phase curves of pure olivine, pure amorphous carbon, and three different mixtures for power-law exponents of the particle size distribution of $2$ and $5$ with $c=0.1~\mu$m.}
\label{fig:ModelParameters3}
\end{figure*}
\end{appendix}
\end{document}